\pdfoutput=1

\IfFileExists{aa.cls}{%
  \documentclass{aa}\def\UsingAAClass{1}%
}{%
  \documentclass[10pt,a4paper,twocolumn]{article}%
}
\ifdefined\UsingAAClass\else
  \usepackage[margin=1.8cm]{geometry}
  \usepackage[round,sort&compress]{natbib}
  \newcommand{\institute}[1]{}
  \newcommand{\offprints}[1]{}
  \newcommand{\titlerunning}[1]{}
  \newcommand{\authorrunning}[1]{}
  \newcommand{\inst}[1]{\textsuperscript{#1}}
  \newcommand{\keywords}[1]{\par\noindent\textbf{Key words.} #1\par}
  \renewcommand{\abstract}[5]{%
    \twocolumn[\maketitle\begin{quote}\small
    \textbf{Context.} #1 \textbf{Aims.} #2 \textbf{Methods.} #3
    \textbf{Results.} #4 \textbf{Conclusions.} #5\end{quote}]}
  \newcommand{\email}[1]{\texttt{#1}}
  \newcommand{\tablefoot}[1]{\par\smallskip\footnotesize #1}
  \newenvironment{acknowledgements}{\par\medskip\noindent\textit{Acknowledgements.}\ }{\par}
\fi

\usepackage[T1]{fontenc}
\usepackage{amsmath,amssymb,bm}
\usepackage{graphicx}
\usepackage{booktabs}
\usepackage{array}
\usepackage{xcolor}
\ifdefined\UsingAAClass\usepackage{hyperref}\else\usepackage{url}\fi
\ifdefined\UsingAAClass
  \makeatletter
  \renewcommand*\aa@institutefont{\small\centering}
  \renewcommand*\aa@datefont{\small\centering}
  \makeatother
\fi

\newcommand{\ExoPlanetLab}{\textsc{ExoPlanetLab}}

\newcommand{\gaia}{\textit{Gaia}}
\newcommand{\mas}{\,\mathrm{mas}}

\newcommand{\yes}{$\bullet$}
\newcommand{\no}{--}

\begin{document}

\title{\ExoPlanetLab: Bayesian orbit fitting of \gaia{}-DR4 epoch astrometry, radial velocities and relative astrometry}
\titlerunning{\ExoPlanetLab}
\author{D.~Hobbs\inst{1}}
\authorrunning{D. Hobbs}
\institute{Lund Observatory, Division of Astrophysics, Department of Physics, \\
Lund University, Box 43, SE-221 00 Lund, Sweden\\
\email{david.hobbs@fysik.lu.se}}
\date{}

\ifdefined\UsingAAClass
\abstract
{\gaia{} Data Release~4 (DR4) provides epoch astrometry from which
thousands of giant planets and many substellar and stellar companions
can be characterised.}
{We present \ExoPlanetLab, an open-source package for Bayesian orbit inference 
from Gaia epoch astrometry, radial velocities (RVs) and relative astrometry, 
validated on simulated and real data.}
{The model orbit is compared with the epoch residuals after removing the part
absorbed by the single-star fit. A Student-$t$ likelihood describes
heavy-tailed data. A robust Keplerian periodogram with bootstrap false-alarm
probabilities gives starting values for a joint fit. Five samplers are
available: tempered ensemble MCMC, two nested samplers and two reversible-jump
samplers. Evidences for a fixed number of companions select that number.
Acceleration and scan-angle models test whether an orbit is warranted.}
{The samplers agree within $0.1\sigma$ on a simulated two-planet system. In
180 injections all fits converge, and all signals with signal-to-noise ratio
$S\geq10$ are detected. The posteriors show no significant bias, but the period
coverage is below nominal: 90\% of the true periods lie within their 95\%
credible intervals. None of 30 companion-free data sets gives a false
detection. On DR4 pre-release data the orbit of Gaia BH3 is recovered within
$1.4\sigma$ of published solutions. The Student-$t$ model brings the
astrometric period and eccentricity of HD\,114762 onto its RV orbit. Eight
simulated relative positions narrow the period's credible interval to 0.27 of
its width, and \texttt{Octofitter} agrees within $0.2\sigma$ on the same data.}
{The code is validated for single sources with DR4-like epoch astrometry, with
or without RVs. Relative astrometry is tested on simulated data only. The main
limits are an uncorrelated noise model and the computing cost, which grows
linearly with the number of measurements and steeply with the number of
companions.}

\ifdefined\UsingAAClass\maketitle\nolinenumbers\fi

\section{Introduction}
\label{sec:intro}

A star with an unseen companion moves on the sky around the system
barycentre. \gaia{} measures that motion one dimension at a time, along the
instantaneous scan direction, with a per-transit precision of tens of
micro-arcseconds for bright stars \citep{GaiaCollaboration2023}. \gaia{} DR3
published orbital solutions for about $1.7\times10^5$ systems
\citep{GaiaArenou2023, Halbwachs2023, Holl2023}. The epoch data behind
them was not released. DR4 publishes the epoch astrometry of all sources.
These data are predicted to yield about $10^6$ astrometric orbits, most of
them of binary stars \citep{ElBadry2024}, and of the order of $10^4$ giant
planets. \citet{Lammers2025} predict $7500\pm2100$ giant planets in DR4 and
$120\,000\pm22\,000$ in DR5. \citet{Perryman2014} predicted $21\,000\pm6000$ and
$70\,000\pm20\,000$, but their criteria also count planets with a poorly
constrained period. With those criteria the DR4 estimate of
\citet{Lammers2025} is about 16\,000. Tens of thousands of brown dwarfs come in
addition \citep{Holl2022}.

Fitting an orbit to \gaia{} epoch astrometry differs from the RV and
direct-imaging problems for which most orbit codes were written. The data are
one-dimensional projections along a scan angle that changes from transit to
transit. The published epoch values are residuals of a single-star fit
(position, parallax, proper motion), which absorbs part of any long-period
orbit. When the period exceeds the time span of the data, only an arc of the
orbit is seen, and many orbits fit it equally well. The posterior is then far
from Gaussian, with long tails in period and separate modes in orbital phase.
Combining the astrometry with ground-based RVs, which may extend decades
beyond the \gaia{} baseline, removes the $M\sin i$ degeneracy. It adds
instrument offsets and jitters, and a sign ambiguity between
$(\Omega,\omega)$ and $(\Omega+\pi,\omega+\pi)$. Several codes now address
parts of this problem (Sect.~\ref{sec:tools}). \ExoPlanetLab{} was written to
study it with one likelihood and several samplers, so that results that depend
on the sampler can be separated from properties of the data.

Sections~\ref{sec:model} and~\ref{sec:inference} describe the model and the
inference pipeline. Section~\ref{sec:sims} validates them on simulations, and
Sect.~\ref{sec:real} on the \gaia{} DR4 pre-release epoch astrometry.
Section~\ref{sec:discussion} summarises the results and discusses the
limitations.

\section{Existing tools}
\label{sec:tools}

Table~\ref{tab:tools} summarises the main publicly available codes that fit
orbits to absolute astrometry, based on their papers and documentation.
\texttt{kima} \citep{Faria2018} has recently been extended with models for
\gaia{} epoch astrometry alone and jointly with RVs, sampled with diffusive
nested sampling with a free number of Keplerians
\citep{Baycroft2026}; it fits the five single-star parameters (optionally seven
or nine, with acceleration terms), an astrometric jitter, an optional
Student-$t$ likelihood, and scan-angle harmonic terms to discriminate
spurious signals, and has been applied to Gaia BH3. \texttt{kepmodel}
\citep{Delisle2022} models astrometric and RV time series with Keplerians,
linear predictors and Gaussian-process noise, and provides a linear
periodogram with analytical false-alarm probabilities. \texttt{Octofitter}
\citep{Thompson2023} fits relative astrometry, RVs, Hipparcos intermediate astrometric
data (IAD), proper-motion anomalies, images and interferometric visibilities, with
Hamiltonian Monte Carlo or parallel-tempered sampling, a variable number of
planets, optional $N$-body integration, and preliminary support for DR4 epoch
astrometry. \texttt{orvara} \citep{Brandt2021orvara} combines RVs, relative
astrometry and the Hipparcos--\gaia{} Catalog of Accelerations
\citep{Brandt2021}; \texttt{orbitize!} \citep{Blunt2020, Blunt2024} targets
directly imaged companions and adds Hipparcos intermediate data and $N$-body
modelling; BINARYS \citep{Leclerc2023} combines Hipparcos and \gaia{} data
with relative astrometry and RVs for binaries. \texttt{gaiamock}
\citep{ElBadry2024} simulates \gaia{} epoch astrometry from the scanning law and
fits it with the DR3 hierarchy of single-star, acceleration and orbital
solutions, and the DR3 orbital pipeline itself is described by
\citet{Holl2023}.

\begin{table*}
\caption{Capabilities of codes that fit orbits to absolute astrometry, from
their papers and documentation (\yes: yes; \no: no).}
\label{tab:tools}
\centering\footnotesize\setlength{\tabcolsep}{2.2pt}
\begin{tabular}{@{}lcccccccc@{}}
\toprule
 & \ExoPlanetLab & \texttt{kima} & \texttt{kepmodel} & \texttt{Octofitter} & \texttt{orvara} & \texttt{orbitize!} & BINARYS & \texttt{gaiamock}\\
\midrule
Language                       & Python & C++/Python & Python & Julia & Python & Python & R/C++ & Python\\
\gaia{} epoch astrometry & \yes & \yes & \yes & \yes$^a$ & \no & \no & \no & \yes$^b$\\
Single-star terms              & projected$^c$ & fitted & fitted & fitted & -- & -- & -- & fitted\\
Hipparcos IAD                  & \yes & \no & \no & \yes & \yes$^d$ & \yes & \yes & \no\\
HGCA / PM anomaly              & \no & \yes$^e$ & \no & \yes & \yes & \yes & \no & \no\\
Relative astrometry            & \yes & \no & \no & \yes & \yes & \yes & \yes & \no\\
Multi-instrument RVs           & \yes & \yes & \yes & \yes & \yes & \yes & \yes & \no\\
Variable number of planets     & \yes & \yes & \no & \yes & \no & \no & \no & \no\\
Evidence ($\ln\mathcal{Z}$)   & \yes & \yes & \no & \yes & \no & \yes & \no & \no\\
Several samplers               & \yes\,(5) & \no & \no$^f$ & \yes & \no & \yes & \no & \no\\
Normalising-flow proposals     & \yes & \no & \no & \no & \no & \no & \no & \no\\
Periodogram + FAP screen       & \yes & \no & \yes & \no & \no & \no & \no & \no\\
Astrometric jitter             & \yes & \yes & \yes & \yes & \no & \no & \no & \no\\
Acceleration models (7/9 par.) & \yes & \yes & \yes$^g$ & \no & \no & \no & \no & \yes\\
Scan-angle harmonic terms      & \yes & \yes & \yes$^g$ & \no & \no & \no & \no & \no\\
Student-$t$ likelihood         & \yes & \yes & \no & \no & \no & \no & \no & \no\\
Correlated noise (GP)          & \no & \yes$^e$ & \yes & \yes & \no & \no & \no & \no\\
$N$-body                       & \no & \no & \no & \yes & \no & \yes & \no & \no\\
\bottomrule
\end{tabular}
\tablefoot{$^a$ Preliminary support. $^b$ Simulation and DR3-style fitting.
$^c$ Removed from the likelihood by the projection $(\mathbf{I}-\mathbf{H})$ and
recovered for each posterior draw (Sect.~\ref{sec:projection}).
$^d$ Through \texttt{htof}. $^e$ In separate models (RVs with the
Hipparcos--\gaia{} proper-motion anomaly; Gaussian processes for RVs), not in
the \gaia{} epoch-astrometry models. $^f$ Maximum likelihood with errors from
the Hessian; no sampler. $^g$ As user-defined linear terms.}
\end{table*}

The main methodological differences between \ExoPlanetLab{} and these packages are threefold.
(i) Before comparing a model orbit with the published residuals, it removes
from the orbit's along-scan signal the part that the single-star fit would
have absorbed, by applying the same linear fit, with the same weights, that
produced the residuals (Sect.~\ref{sec:projection}). (ii) The same likelihood can be sampled with five different
algorithms, which exposes sampler-dependent failures and has been essential
for the multi-planet case (Sect.~\ref{sec:sims}). (iii) It includes a
Keplerian periodogram with bootstrap false-alarm probabilities and a
cross-refinement step for commensurate periods (in a ratio of small integers,
such as 2:1). The code provides an opt-in Hipparcos IAD mode (with warnings), validated
against the Hipparcos catalogue, injections and \texttt{Octofitter} on
HD\,114762 (described in the user guide that comes with the code; see
Code availability). By default, however,
the pipeline operates on \gaia{} epoch astrometry, optionally joint with RVs
and relative (imaging) astrometry (Sect.~\ref{sec:relast}): it avoids
combining multi-catalogue proper-motion anomalies (such as the HGCA), whose
cross-mission systematics are complex, so that all orbital constraints derive
from observational data whose error models can be evaluated directly.
Acceleration solutions are included (Sect.~\ref{sec:accel}): they decide whether an orbit is
warranted when the period exceeds the time span of the data.

\section{Model}
\label{sec:model}

\subsection{Astrometric orbit}
\label{sec:orbit}

Each orbiting companion $k$ is described by seven parameters $(P_k, e_k, T_{0,k}, A_k, B_k, F_k, G_k)$.
With $M = 2\pi(t-T_0)/P = E - e\sin E$, $X=\cos E-e$ and
$Y=\sqrt{1-e^2}\sin E$, the photocentre displacement
$\mathbf p(t)=(\Delta\alpha^*,\Delta\delta)$ is
\begin{equation}
  \Delta\delta = AX + FY,\qquad \Delta\alpha^* = BX + GY,
\end{equation}
in the standard Thiele--Innes convention (orbital-plane $x$ axis to the
north), and the along-scan signal at scan angle $\psi_i$ is
\begin{equation}
  s_i = \Delta\alpha^*_i\sin\psi_i + \Delta\delta_i\cos\psi_i,
\label{eq:al}
\end{equation}
and with several companions the signals of the individual companions add.
The Campbell elements $(a_0, i, \Omega, \omega)$
follow from the constants \citep{Green1985}. Because the constants enter
linearly, no trigonometry of the orientation angles is evaluated in the
likelihood, and the periodogram of Sect.~\ref{sec:screen} can solve for them
in closed form.

\subsection{Projection through the single-star fit}
\label{sec:projection}

The DR4 epoch values are not the measured along-scan positions themselves
but the residuals, observed minus computed (O$-$C), of the single-star
astrometric solution. Let $\mathbf{d}$ be the $N_{\rm obs}$ measured
along-scan positions, $\mathbf{D}$ the $N_{\rm obs}\times p$ design matrix of
the single-star model (position, parallax, proper motion and a pseudo-colour
term with prior precision $\bm\Lambda$; it is usually called $\mathbf{A}$ in
\gaia{} papers, a symbol we keep for a Thiele--Innes constant), and
$\mathbf{W}=\mathrm{diag}(\sigma_i^{-2})$ the weights. The pipeline estimates
the single-star parameters by weighted least squares,
\begin{equation}
  \bm\beta=\mathbf{N}^{-1}\mathbf{D}^\top\mathbf{W}\mathbf{d},\qquad
  \mathbf{N}=\mathbf{D}^\top\mathbf{W}\mathbf{D}+\bm\Lambda,
\end{equation}
and publishes the residuals
\begin{equation}
  \mathbf{r}=\mathbf{d}-\mathbf{D}\bm\beta\qquad(\mathrm{O}-\mathrm{C}).
\end{equation}

If the star has a companion, $\mathbf{d}=\mathbf{D}\bm\beta_0+\mathbf{s}(\bm\theta)+\mathbf{n}$,
with $\bm\beta_0$ the true single-star parameters, $\mathbf{s}$ the along-scan orbit signal of
Sect.~\ref{sec:orbit} and $\mathbf{n}$ the noise. The single-star solution knows nothing of the orbit
and absorbs into $\bm\beta$ whatever part of it resembles a change of
position, parallax or proper motion. To see what is left, substitute the
solution for $\bm\beta$ into $\mathbf{r}$:
\begin{equation}
  \mathbf{r}=(\mathbf{I}-\mathbf{H})\,\mathbf{d},\qquad
  \mathbf{H}=\mathbf{D}\,\mathbf{N}^{-1}\mathbf{D}^\top\mathbf{W}.
\end{equation}
Applied to any vector of along-scan values, $\mathbf{H}$ returns the
best-fitting single-star curve (it is the hat matrix of the fit;
\citealt{Hoaglin1978}), so $(\mathbf{I}-\mathbf{H})$ returns the O$-$C of a
single-star fit to that vector. A pure single-star signal is fitted exactly,
$(\mathbf{I}-\mathbf{H})\mathbf{D}\bm\beta_0=0$ (for $\bm\Lambda=0$; the prior is
negligible in practice, Sect.~\ref{sec:real}), and therefore
\begin{equation}
  \mathbf{r}=(\mathbf{I}-\mathbf{H})\,\mathbf{s}(\bm\theta)+(\mathbf{I}-\mathbf{H})\,\mathbf{n}.
\end{equation}
The model to compare with the published residuals is not the orbit
$\mathbf{s}(\bm\theta)$ but $(\mathbf{I}-\mathbf{H})\,\mathbf{s}(\bm\theta)$, i.e.,
the part of the orbit that a single star can mimic has been removed from the
data and must be removed from the model in the same way. In \gaia{} terms,
$(\mathbf{I}-\mathbf{H})\,\mathbf{s}$ is what one obtains by running the
single-star solution, with the pipeline's weights, on the noise-free orbit
signal and taking its O$-$C. For a period long compared with the time span of
the data this removes most of the orbit. With the weights absorbed into the
design matrix, $\tilde{\mathbf{D}}=\mathbf{W}^{1/2}\mathbf{D}$, the
corresponding symmetric projector is
$\tilde{\mathbf{D}}\mathbf{N}^{-1}\tilde{\mathbf{D}}^\top$; its diagonal gives
the fraction of each measurement absorbed by the single-star fit, and its
trace is $p$. \ExoPlanetLab{} rebuilds
$\mathbf{H}$ from the processing, verifies that it reproduces the delivered
residuals to $<10^{-6}\mas$, and applies it to every model compared with the
data. The projection is required. Without it, a published orbit fits the
published residuals worse than no orbit at all (Sect.~\ref{sec:real}).

\paragraph{Projected versus fitted single-star terms.} Other codes (e.g.\
\texttt{kima}) instead fit the single-star parameters $\bm\beta$ together with
the orbit, modelling $\mathbf{d}=\mathbf{D}\bm\beta+\mathbf{s}(\bm\theta)+\mathbf{n}$.
With Gaussian noise and a flat prior on $\bm\beta$ the two approaches give the
same orbital posterior. Writing
$\lVert\mathbf{x}\rVert^2_{\mathbf{W}}=\mathbf{x}^\top\mathbf{W}\mathbf{x}=\sum_i x_i^2/\sigma_i^2$
for the $\chi^2$ of a residual vector,
\begin{equation}
  \bigl\lVert\mathbf{r}-(\mathbf{I}-\mathbf{H})\mathbf{s}\bigr\rVert^2_{\mathbf{W}}
  = \min_{\bm\beta}\bigl\lVert\mathbf{d}-\mathbf{s}-\mathbf{D}\bm\beta\bigr\rVert^2_{\mathbf{W}}:
\end{equation}
the left side is our $\chi^2$, the right side that of a joint fit at the best
single-star parameters, and both equal
$\lVert(\mathbf{I}-\mathbf{H})(\mathbf{d}-\mathbf{s})\rVert^2_{\mathbf{W}}$,
because the minimising $\bm\beta$ is the single-star solution for the data
$\mathbf{d}-\mathbf{s}$. This is the Frisch--Waugh--Lovell theorem of linear
regression \citep{FrischWaugh1933,Lovell1963}. Removing the linear parameters
of a nonlinear fit in this way is also known as variable projection
\citep{GolubPereyra1973}. Our $\chi^2$ is therefore the fitted one profiled
over $\bm\beta$, and marginalising over
$\bm\beta$ instead adds only $-\frac12\ln\det\mathbf{N}$ to the
log-likelihood, a constant because $\mathbf{N}$ does not depend on
$\bm\theta$. 

The equivalence is exact for Gaussian errors without a prior on
$\bm\beta$. The weak pseudo-colour prior makes it approximate, negligibly so in
practice (Sect.~\ref{sec:real}), and so does a fitted jitter
(Sect.~\ref{sec:jitter}), at the level of $p/N_{\rm obs}\approx1\%$ in the
jitter variance, because $\mathbf{H}$ keeps the pipeline's weights. The
differences between the two approaches are practical. Projection samples seven
parameters per companion rather than $7N_{\rm pl}+p$, reuses the pipeline's
own weights and prior, and needs no priors on $\bm\beta$. It does not report
$\bm\beta$, the orbit-corrected parallax and proper motion, but these can be
recovered for each posterior sample as
$\mathbf{N}^{-1}\mathbf{D}^\top\mathbf{W}(\mathbf{d}-\mathbf{s})$; and it
cannot take an informative prior on the parallax. Finally, the projection
assumes Gaussian errors: a least-squares fit is pulled by outliers, and
$(\mathbf{I}-\mathbf{H})$ spreads their effect over the other epochs. For
heavy-tailed noise \ExoPlanetLab{} therefore refits $\bm\beta$ robustly at
every likelihood evaluation (Sect.~\ref{sec:jitter}).

\subsection{Radial velocities}
\label{sec:rv}

The RV model is
$V_{r,j} = \sum_k K_k[\cos(\nu_k+\omega_k)+e_k\cos\omega_k] + \gamma_{\iota(j)}$,
with $\nu_k$ the true anomaly of companion $k$ and $\omega_k$ derived from
the constants, so the RVs share $(P,e,T_0,\omega)$ with the astrometry and add
only $K_k$ and one offset $\gamma_\iota$ and jitter $s_\iota$ per instrument $\iota$. The constants determine
$(\Omega,\omega)$ only up to $(\Omega+\pi,\omega+\pi)$; $\omega$ is taken on
the branch with $0\le\Omega<\pi$ and $K$ is allowed to be negative, which
selects the other node. Reported values are converted to $K>0$, moving both
angles by $\pi$ when $K<0$, so that with RVs the node is determined over the
full circle. The bounds on $K$, $\gamma$
and the jitter scale with the range of the loaded RVs, so that absolute
velocities of stellar-mass companions are handled without clipping. RV times
must declare their zero-point, and are shifted onto the astrometric one.

For a dark companion the photocentre orbit is the orbit of the star, so $K$
follows from the astrometric orbit and the parallax. $K$ can optionally be tied
to the orbit in this way, either in the fit or afterwards by importance
sampling of the posterior draws of a fit with $K$ free; the two give the same
result.

\subsection{Relative astrometry}
\label{sec:relast}

A companion resolved by imaging, interferometry or speckle observations is
measured relative to its host. With $f_{\rm mass}=M_2/(M_1+M_2)$ the mass fraction and
$f_{\rm flux}=L_2/(L_1+L_2)$ the flux fraction, the host moves about the barycentre by
$-f_{\rm mass}\,\mathbf r$ and the unresolved photocentre by $-(f_{\rm mass}-f_{\rm flux})\,\mathbf r$,
where $\mathbf r$ is the companion's position relative to the host, so that
\begin{equation}
\mathbf r(t) = -s\,\mathbf p(t), \qquad
s=\frac{a_{\rm rel}}{a_0}=\frac{1}{f_{\rm mass}-f_{\rm flux}}>1 ,
\label{eq:relast}
\end{equation}
with $\mathbf p(t)$ the photocentre displacement of Sect.~\ref{sec:orbit} and
$a_{\rm rel}$ and $a_0$ the angular semi-major axes of the relative and
photocentre orbits. The
relative positions thus share all orbital elements with the \gaia{} data and
add one parameter per imaged companion, $s$, with a log-uniform prior. Their
likelihood is a two-dimensional Gaussian per epoch with the quoted errors and
correlation, to which a fitted jitter can be added in quadrature; the reflex
motion of the host due to other companions is neglected.

The scale $s$ separates the two masses: Eq.~\eqref{eq:relast} is the
photocentre relation used for the \gaia{} DR3 astrometric binaries
(Eq.~14 of \citealt{Halbwachs2023}; Eq.~2 of \citealt{GaiaArenou2023}), written
with $f_{\rm mass}$ and $f_{\rm flux}$. With the parallax, the relative semi-major axis in
au is $a_{\rm rel}/\varpi=s\,a_0/\varpi$, and Kepler's third law gives the
total mass $M_1+M_2=(a_{\rm rel}/\varpi)^3/P^2$ (au, yr, $M_\odot$) without a
stellar model. Solving Eq.~\eqref{eq:relast} for the mass fraction gives
$f_{\rm mass}=1/s+f_{\rm flux}$, and the individual masses follow as
$M_2=f_{\rm mass}(M_1+M_2)$ and $M_1=(1-f_{\rm mass})(M_1+M_2)$. For a dark
companion $f_{\rm flux}=0$, so $M_2=(M_1+M_2)/s$; for a luminous one
$f_{\rm flux}$ must come from photometry, from the measured flux ratio
$L_2/L_1$ in the $G$ band, and it enters only this mass calculation in which $s$ is free.

\subsection{Likelihood and priors}

For Gaussian noise the astrometric and RV terms are independent:
\begin{multline}
\ln\mathcal{L}_{\rm Gauss} = -\tfrac12\sum_i\Bigl[\frac{(r_i-m_i)^2}{\sigma_i^2+s_{\rm ast}^2} +\ln(\sigma_i^2+s_{\rm ast}^2)\Bigr] \\
 -\tfrac12\sum_j\Bigl[\frac{(v_j-V_{r,j})^2}{\sigma_j^2+s_{\iota(j)}^2} +\ln(\sigma_j^2+s_{\iota(j)}^2)\Bigr] + \mathrm{const}.
\label{eq:gauss_like}
\end{multline}
The first sum runs over the $N_{\rm obs}$ along-scan measurements: $r_i$ is the
published residual and $m_i=[(\mathbf{I}-\mathbf{H})\,\mathbf{s}(\bm\theta)]_i$
the projected orbit model (Sect.~\ref{sec:projection}), $\sigma_i$ the formal
uncertainty of measurement $i$, and $s_{\rm ast}$ the astrometric jitter
(Sect.~\ref{sec:jitter}), added in quadrature. The second sum runs over the
$N_{\rm RV}$ radial velocities: $v_j$ is the measured velocity with
uncertainty $\sigma_j$, $V_{r,j}$ the model of Sect.~\ref{sec:rv}, and $\iota(j)$ the
instrument that made measurement $j$, with jitter $s_{\iota(j)}$. The constant is
$-\tfrac12(N_{\rm obs}+N_{\rm RV})\ln2\pi$. The logarithmic terms normalise
each Gaussian and matter only when a jitter is fitted: with all jitters fixed
they are constant, and Eq.~\eqref{eq:gauss_like} reduces to
$-\tfrac12\chi^2$, the weighted least-squares $\chi^2$ of
Sect.~\ref{sec:projection} plus its RV counterpart.
Priors are log-uniform on $P$ between 1\,d and twice the nominal 10-yr mission
(7305\,d, about four DR4 baselines; a larger upper bound let samplers settle
on spurious long-period, high-eccentricity fits), uniform on $e$, on $T_0$ over
the baseline extended by one maximal period on either side, isotropic in
orientation with a log-uniform $a_0$ (below), uniform on $K$
and $\gamma$, log-uniform on the jitters (an RV instrument with fewer than
three points has no free jitter: with one point its offset reproduces the
measurement exactly, so the data say nothing about the jitter, and the
posterior in offset and jitter becomes a funnel, the offset pinned at small
jitter and free at large jitter, which nested samplers cannot explore; the
jitter is then held at 0.01\,m\,s$^{-1}$), and Gaussian
on the stellar mass and
parallax when their external uncertainties are given (these enter only the
mass estimates). Multi-planet fits impose $P_1<P_2<\dots$. Uniform priors on
the Thiele--Innes constants imply $p(a_0,i,\Omega,\omega)\propto a_0^3\sin^3 i$,
the Jacobian of the transformation from $(a_0,i,\Omega,\omega)$ to $(A,B,F,G)$,
which favours large, edge-on orbits; \citet{Baycroft2026} find that such priors
can distort Bayes factors for low-amplitude signals. By default the prior is
therefore isotropic ($\cos i$, $\Omega$, $\omega$ uniform) with a log-uniform $a_0$
($\ge10^{-3}\mas$), i.e.\ $p(a_0,i,\Omega,\omega)\propto a_0^{-1}\sin i$.
Divided by the Jacobian, this is a factor $a_0^{-4}\sin^{-2}i$ in
Thiele--Innes space, within $|A|,|B|,|F|,|G|<500\mas$; $\sin i$ is floored at
$\sin2^\circ$ to keep the factor finite for face-on orbits. The
uniform Thiele--Innes prior remains an option; it is the only one for the RJ
samplers and, outside evidence runs (Sect.~\ref{sec:evidence}), for
\texttt{dynesty}, whose prior transforms assume it. It should not be used for
weak signals (Sect.~\ref{sec:priorsens}). For the period of an orbit
longer than the baseline the posterior depends on the upper bound of the period
prior (Sect.~\ref{sec:injrec}).

\subsection{Astrometric jitter}
\label{sec:jitter}

Real epoch astrometry is not fully described by its formal errors
(Sect.~\ref{sec:real}). An astrometric jitter $s_{\rm ast}$, one parameter per
source, is therefore added in quadrature to every along-scan error,
$\sigma_i^2\to\sigma_i^2+s_{\rm ast}^2$, and fitted with the orbit (log-uniform
prior from $10^{-3}$\,mas to the larger of 1\,mas and three times the scatter of
the data); it is identifiable because the likelihood keeps its $\ln\sigma_i$
term. It measures how much noise the formal errors miss, whatever its origin.

For the DR4 pre-release data a single white jitter is the wrong shape. With the
formal errors, the orbit fits give $\chi^2/\mathrm{dof} = 1.4$, $3.1$ and $9.9$
for Gaia-4, Gaia BH3 and HD\,114762. The scatter of the CCD measurements of one
transit about their mean gives almost the same values: $1.4$, $3.1$ and $9.8$.
The whole excess is therefore already present within a transit, and it is not a
per-transit (attitude) term. The core of the distribution follows the formal
errors. The excess comes from a few per cent of outliers, up to $46\sigma$. The default model for epoch data is therefore a
Student-$t$ distribution per CCD measurement \citep{Lange1989}, $t_\nu(x; f^2\sigma_i^2)$, of
scale $f\sigma_i$ and with two parameters per source: the number of degrees of freedom $\nu$, which
sets the weight of the tails (small $\nu$ gives heavy tails, and
$\nu\to\infty$ the Gaussian), and a factor $f$ that scales the formal errors
$\sigma_i$ of the core. Both are fitted, with log-uniform priors on $[1,100]$
and $[0.1,10]$. Because outliers also corrupt a linear least-squares
single-star fit, the robust likelihood avoids using the published residuals
and instead uses the raw along-scan positions,
$d_i = s_i(\bm\theta) + (\mathbf{D}\bm\beta)_i + \varepsilon_i$, with
$s_i(\bm\theta)$ the unprojected orbit signal of Sect.~\ref{sec:orbit} and
noise $\varepsilon_i$ distributed as $t_\nu(\varepsilon_i; f^2\sigma_i^2)$. The single-star
terms are refitted for every likelihood evaluation, by iteratively reweighted
least squares (IRLS): starting from the least-squares solution, each
measurement is weighted by $w_i = (\nu+1)/(\nu+r_i^2/f^2\sigma_i^2)$, which
down-weights large residuals, and the weighted fit is repeated until the best
fit $\hat{\bm\beta}$ converges. The weights follow from the likelihood. Setting
the derivative of $\sum_i\ln t_\nu$ with respect to $\bm\beta$ to zero gives the
normal equations of a weighted least-squares fit,
$\sum_i w_i r_i\mathbf{D}_i/f^2\sigma_i^2=\mathbf{0}$, with $\mathbf{D}_i$ the
$i$-th row of $\mathbf{D}$. The weights depend on the residuals, hence the
iteration \citep{HollandWelsch1977}. This is the maximum-likelihood fit of a
regression with $t$-distributed errors \citep{Lange1989}. Here
$r_i = d_i - s_i(\bm\theta) - (\mathbf{D}\hat{\bm\beta})_i$ are the residuals
of this robust fit. Marginalising over $\bm\beta$ with a Laplace
approximation \citep{TierneyKadane1986} about $\hat{\bm\beta}$, in which the
Fisher information of the $t$-distribution \citep{Lange1989}, $c/f^2\sigma_i^2$ per measurement, takes the place of the
Gaussian weights $1/\sigma_i^2$, gives
\begin{equation}
\begin{split}
\ln\mathcal{L}_{\rm robust} = {}&\sum_i \ln t_\nu(r_i; f^2\sigma_i^2) - \tfrac12\hat{\bm\beta}^\top \bm\Lambda \hat{\bm\beta} \\
 &-\tfrac12\ln\det\bigl(\mathbf{D}^\top \mathrm{diag}(c/f^2\sigma_i^2)\mathbf{D} + \bm\Lambda\bigr) + \mathrm{const},
\end{split}
\label{eq:robust_like}
\end{equation}
with $c=(\nu+1)/(\nu+3)$, and $\bm\Lambda$ the pseudo-colour prior of
Sect.~\ref{sec:projection}. With RVs, the RV sum of Eq.~\eqref{eq:gauss_like} is added to
Eq.~\eqref{eq:robust_like} unchanged; $\nu$ and $f$ enter only the astrometric term. As $\nu\to\infty$ and $f\to1$, Eq.~\eqref{eq:robust_like} strictly converges to the
exact Gaussian marginal likelihood over single-star terms of
Sect.~\ref{sec:projection}. The fitted tails are
$\nu = 8.5$, $3.0$ and $2.2$ with $f = 1.02$, $0.99$ and $0.84$; on simulated Gaussian data
$\nu$ goes to large values, $f\approx1$ and the orbit is unchanged, the widths growing by
2\%. The Gaussian model with jitter remains available.

\subsection{Acceleration and scan-angle terms}
\label{sec:accel}

A companion with a period much longer than the data shows only the curvature
of its orbit: a constant acceleration of the proper motion, or an acceleration
and its rate of change (jerk). Such terms are part of the core astrometric
solution (AGIS) used for the DR4 epoch astrometry, and in DR3 they gave the
seven- and nine-parameter acceleration solutions of the non-single-star
processing \citep{Halbwachs2023}. We add them as linear terms of the single-star model,
\begin{equation}
  \Delta\alpha^* = \tfrac12 a_\alpha\tau^2 + \tfrac16 j_\alpha\tau^3,\qquad
  \Delta\delta = \tfrac12 a_\delta\tau^2 + \tfrac16 j_\delta\tau^3,
\end{equation}
with $\tau$ the time in years from the reference epoch of the epoch data,
$(a_\alpha,a_\delta)$ the acceleration (mas\,yr$^{-2}$) and
$(j_\alpha,j_\delta)$ the jerk (mas\,yr$^{-3}$) in the two coordinates; the
seven-parameter model has the acceleration only. In the along-scan direction
these displacements are projected as in Eq.~\eqref{eq:al}, and the fitted
parameters are the acceleration and jerk themselves. Two further models test
a different explanation: an along-scan time independent shift that depends on 
the direction of the scan, $\psi$,
\begin{equation}
  \Delta_{\rm AL} = \sum_n \bigl(u_n\cos n\psi + v_n\sin n\psi\bigr),
  \label{eq:harm}
\end{equation}
with harmonic amplitudes $(u_n, v_n)$ in mas, fitted directly, and $n=2$ or
$n=2,3$. The $n=1$ harmonic, $u_1\cos\psi + v_1\sin\psi$, is a fixed offset in
$\delta$ and $\alpha^*$ and is already in the single-star model. Such a signal
arises from source structure or calibration errors, which the scanning law can
turn into an apparent orbit \citep{Holl2023}. It is an alternative to an orbit
or an acceleration and not a reparametrisation of them. Being linear, all these terms are fitted with
the single-star terms at every likelihood evaluation and cost no sampled
parameter. The evidence of these models needs a proper prior on the terms,
which is built in two levels. Each pair, $(a_\alpha,a_\delta)$,
$(j_\alpha,j_\delta)$ or $(u_n,v_n)$, has a Gaussian prior with zero mean and
standard deviation $\sigma_{\rm p}$ on each component,
$\mathcal N(0,\sigma_{\rm p}^2)$; $\sigma_{\rm p}$ sets the typical size of the
terms and is itself not fixed but log-uniform over five decades, $10^{-3}$ to
$100$ in the unit of the terms (mas\,yr$^{-2}$, mas\,yr$^{-3}$ or mas). The
mixture of Gaussians over $\sigma_{\rm p}$ \citep{AndrewsMallows1974} is isotropic in direction and close
to log-uniform in amplitude, like the $a_0$ prior of an orbit. The first level
can be integrated without sampling. For a fixed $\sigma_{\rm p}$ both the
likelihood and the prior are Gaussian in the linear terms (the likelihood
exactly for Gaussian noise, and in the Laplace approximation of
Sect.~\ref{sec:jitter} for Student-$t$ noise), so the integral of their product
over the terms has the closed form \citep{MacKay1992}
\begin{equation}
  \ln\mathcal{Z}(\sigma_{\rm p}) = \ln\mathcal{L}_{\rm flat}
   - \tfrac12\ln\det(\mathbf C+\mathbf S)
   - \tfrac12\mathbf b^\top(\mathbf C+\mathbf S)^{-1}\mathbf b,
\end{equation}
where $\mathcal{L}_{\rm flat}$ is the likelihood with the new terms added to
the single-star model and given a flat prior (Sects.~\ref{sec:projection}
and~\ref{sec:jitter}), $\mathbf b$ and $\mathbf C$ are the best fit of the new
terms and their covariance under that flat prior, and
$\mathbf S=\sigma_{\rm p}^2\mathbf I$ is the covariance of the Gaussian prior.
With $\mathcal{L}_{\rm flat}$ normalised as in Eq.~\eqref{eq:robust_like}, the
factors of $2\pi$ of the two Gaussian integrals cancel.
The last two terms are the price of the prior. The quadratic term penalises
fitted terms that are large compared with $\sigma_{\rm p}$. The determinant
term is an Occam penalty: it grows with how much more precisely the data fix
a term than the prior does, and it vanishes for a term that the data do not
constrain. The second level,
the average of $\mathcal{Z}(\sigma_{\rm p})$ over the log-uniform prior on
$\sigma_{\rm p}$ (one scale per pair), is computed numerically on a grid. The
models are used in two ways:
alone, their evidences are direct integrals (a few seconds per source) that an
orbit has to beat; inside an orbit fit, they absorb the curvature of a wider
companion, or a scan-angle signal, without sampling it. With RVs a linear
velocity trend common to all instruments can be integrated in the same way.

\section{Inference}
\label{sec:inference}

\subsection{Signal screen}
\label{sec:screen}

A time-only periodogram is invalid for scan-projected data, and a
circular-orbit approximation places the maximum at the wrong period for
eccentric orbits. The screen therefore profiles $e$ and $T_0$ on a grid at each
trial period and, since the signal is linear in the Thiele--Innes constants,
solves for them by weighted least squares, with the same projection
$(\mathbf{I}-\mathbf{H})$ as the orbit model (Sect.~\ref{sec:projection}). The significance of a peak is tested with a
permutation bootstrap: the residuals are shuffled among the epochs, keeping
the times and scan angles, which destroys any orbital signal but keeps the
noise and the sampling, and the highest peak of each shuffled periodogram
gives its distribution for noise alone. The false-alarm probability (FAP) is
the fraction of 200 shuffles whose highest peak reaches the observed one, and
a peak is accepted if it is below 1\%. Peaks are further vetted by a bounded
refinement with a failure sentinel and a rejection of fits pinned to the
period boundary. A Gauss--Seidel cross-refinement then
re-fits each candidate against the data with only the other candidates
subtracted, removing the duplicate and harmonic peaks that sequential
subtraction produces at commensurate periods. The screen provides the signal
count and the starting point; no fit is initialised from a published orbit.

Least squares is the right statistic only for Gaussian noise; in the
heavy-tailed noise of the epoch data a few outliers dominate $\chi^2$. With the
Student-$t$ noise model the screen therefore first fits $t_\nu(0,f^2\sigma_i^2)$
to the residuals $y_i$ and works on their score,
$y'_i=f\sigma_i\,\phi(y_i/f\sigma_i)$ with $\phi(r)=(\nu+1)r/(\nu+r^2)$, the
locally most powerful statistic for a small linear signal in $t$ noise; the
periodogram and the permutation bootstrap are unchanged, and for Gaussian
residuals $y'=y$ (Sect.~\ref{sec:detection}).

\subsection{Peel and joint fit}
\label{sec:peel}

For $N_{\rm pl}>1$ planets, single-planet ``peel'' fits on progressively cleaned residuals
provide warm starts. At commensurate periods the peels are biased, because
each one absorbs part of the signal of the others (Sect.~\ref{sec:sims}).
Every backend therefore depends on a joint fit of all planets. This is an
affine-invariant ensemble MCMC \citep{GoodmanWeare2010, ForemanMackey2013},
parallel-tempered with four temperatures. With RVs, tempering is needed
because RVs taken years after the astrometry make the posterior multimodal in
phase. Without RVs, it is what makes weak signals, near-circular orbits and
periods beyond the baseline converge (Sect.~\ref{sec:injrec}).

Tempering does not free the fit from its starting point when the modes are
separated in period, so the start is searched for
(Sect.~\ref{sec:multi}). For one orbit the candidates are the screen's
detections and the five highest distinct peaks of the first periodogram; each
is refitted as a single orbit to the full data and they are ranked by the
likelihood of the noise model of the fit. For two or more orbits a beam search
builds the combinations up orbit by orbit: every kept combination is extended
by the three highest distinct peaks of the robust periodogram of its residual,
its orbits are refitted in turn against each other, and the four combinations
with the highest likelihood are kept. The fit starts from the best-ranked
candidate, or the peel start if that is better, and can be repeated from the
others, the highest evidence being kept.

\subsection{Samplers}

\emph{emcee}. The joint fit of Sect.~\ref{sec:peel} is itself the
\emph{emcee} result.

\emph{dynesty} \citep{Speagle2020}. A cold nested run over the full period
prior cannot find the orbit. \emph{dynesty} therefore samples a flat box of
$\pm10\sigma$ around the joint fit in the orbital parameters. The other
parameters keep their priors. The returned $\ln\mathcal{Z}$ is relative to this
box. In evidence runs it is given for the full prior
(Sect.~\ref{sec:evidence}).

\emph{nessai} \citep{Williams2021}. Normalising-flow nested sampling is used
for the peels. Each peel samples a box around its screen candidate: $P\pm20\%$,
$T_0$ over one period, and each Thiele--Innes constant within
$\pm(3\max(|A|,|B|,|F|,|G|)+1\mas)$ of the candidate value. With this box a
peel takes $4\times10^4$ likelihood calls. Over the full prior it took
$2.5\times10^6$ calls or converged on a wrong mode. The joint stage then runs
with \emph{emcee}.

\emph{rjmcmc} and \emph{rjflow}. These two reversible-jump samplers use
\texttt{eryn} \citep{Karnesis2023}. They sample the number of planets
$N_{\rm pl}$ together with the orbits. Both start from an \emph{emcee} joint
fit at the detected number of planets. Births are proposed from its
posterior and from the strongest peak of the residual periodogram. Within a
model both samplers mix differential-evolution and adaptive Gaussian moves.
\emph{rjmcmc} adds a KDE move and \emph{rjflow} a RealNVP flow, both built
from the \emph{emcee} posterior. Both run with parallel tempering. Neither
supports RVs.

\subsection{Model comparison}
\label{sec:evidence}

The reversible-jump samplers accept almost no trans-dimensional moves on these
problems, so the number of companions is chosen from the evidences
$\mathcal{Z}$ of fixed-$N_{\rm pl}$ fits: stepping up from $N_{\rm pl}=0$, a
companion is added when the Bayes factor
$B(N_{\rm pl}\ \mathrm{vs}\ N_{\rm pl}-1)=\mathcal{Z}_{N_{\rm pl}}/\mathcal{Z}_{N_{\rm pl}-1}$
satisfies $\ln B>5$
(odds of about 150:1). For $N_{\rm pl}=0$ and for the acceleration and scan-angle
models only the noise parameters remain, and the evidence is integrated
directly. For orbits it comes from the tempered fit with 16 temperatures down
to the prior, each rung's evidence ratio being estimated by bridge sampling
\citep{Bennett1976} from the samples of both its temperatures; the error is the
scatter of consecutive blocks of the chain.

A strong signal switches on over a narrow range of temperature, and the chains
there equilibrate far more slowly than the cold chain, whose autocorrelation
time normally ends the run. A temperature ladder frozen after the burn-in can
then leave a gap at the transition, and $\ln\mathcal{Z}$ keeps changing with
the length of the run (Sect.~\ref{sec:real}). An evidence run therefore stops only when the two
halves of its production chain give the same $\ln\mathcal{Z}$ within 0.5; when
they do not, the ladder adapts again and the evidence restarts, the cold chain
keeping every step. A run still unstable at its step limit is flagged.

The nested samplers give an independent value. They search a box around the
posterior; in evidence runs they integrate the likelihood times the full
normalised prior over that box (including the isotropic orientation prior and,
for $N_{\rm pl}\ge2$, the period ordering), with the periastron time sampled as
an orbital phase, because the likelihood repeats every period in $T_0$ and a box
around one copy would miss the others. With RVs the nested values are not yet
validated, and the tempered evidence is used (Sect.~\ref{sec:real}).

\subsection{Diagnostics}

Every run records its random seed, settings and package versions, so that it
can be repeated exactly, and reports the integrated autocorrelation times against a $100\tau$
criterion (at which the joint fit stops, except in evidence runs), the joint residual RMS in units of the formal errors, the $T_0$
folded to the first observed cycle together with its raw value, and the
companion mass from the orbit and, with RVs, from $K$ and the astrometric
inclination.

\section{Validation on simulated data}
\label{sec:sims}

\subsection{A commensurate two-planet system}

Two planets with $P=1450$ and $2900$\,d, $e=0.54$ and $0.20$ and
$a_0\approx10.2$ and $4.2\mas$ (SNR 53 and 19) were injected into 140
along-scan epochs over 9.9\,yr with $0.1\mas$ white noise; the RV runs add 50
epochs with $10$\,m\,s$^{-1}$ noise and $K=500$, $200$\,m\,s$^{-1}$. The 2:1
ratio makes single-planet fits biased, and the outer planet is observed for
only 1.2 orbits.

Without RVs, all five samplers (\texttt{emcee}, \texttt{nessai},
\texttt{dynesty} and the two RJ samplers, whose parameters come from the joint
MCMC) agree to within $0.1\sigma$ in every parameter, with a joint residual of
$1.04$--$1.05$ times the noise. All eight runs were made side by side with the
jitter tests of Sect.~\ref{sec:simjitter} and the DR4 runs of
Sect.~\ref{sec:real}, one CPU core each and up to nine at a time, on a 2021
MacBook Pro with an Apple M1 Max (10 cores: 8 performance and 2 efficiency;
64\,GB), in 45\,min in total. Each run took 8--19\,min for
\texttt{emcee}, \texttt{nessai} and \texttt{dynesty} and 29--31\,min for the RJ
samplers without RVs, and 9--13\,min with RVs; the periodogram screen takes
0.5--1.8\,min, and the tempered joint sampling 5--13\,min without RVs and
3--9\,min with them, since it stops once the chain exceeds $100\tau$. The outer planet's $T_0$ and $(A,F,G)$ lie along a nearly
degenerate ridge, and its median sits $2$--$3\sigma$ from the injected values
in every sampler. With RVs, \texttt{emcee}, \texttt{dynesty} and
\texttt{nessai} again agree, every offset falls to $\le1.1\sigma$ and the
$T_0$ width halves (Table~\ref{tab:sim}). The RJ samplers return $N_{\rm pl}=2$ but
accepted no trans-dimensional move. Figure~\ref{fig:corner_sim} shows the
ridge and its removal by the RVs.

\begin{table}
\caption{Outer planet of the simulated 2:1 system: posterior median and
offset from the injected value, without and with RVs.}
\label{tab:sim}
\centering\footnotesize\setlength{\tabcolsep}{3pt}
\begin{tabular}{@{}lrrrrr@{}}
\toprule
 & true & \multicolumn{2}{c}{astrometry} & \multicolumn{2}{c}{+ RV}\\
\midrule
$P$ (d)   & 2900 & $2898.6\pm5.2$ & $-0.3\sigma$ & $2901.7\pm4.4$ & $+0.4\sigma$\\
$T_0$ (d) & 800  & $911\pm47$     & $+2.3\sigma$ & $828\pm28$     & $+1.0\sigma$\\
$A$ (mas) & 1.8  & $1.54\pm0.14$  & $-1.9\sigma$ & $1.75\pm0.07$  & $-0.7\sigma$\\
$B$ (mas) & 3.5  & $3.66\pm0.04$  & $+3.1\sigma$ & $3.56\pm0.06$  & $+0.8\sigma$\\
$F$ (mas) & $-1.0$ & $-1.41\pm0.16$ & $-2.4\sigma$ & $-1.12\pm0.11$ & $-1.1\sigma$\\
$G$ (mas) & 1.2  & $0.36\pm0.38$  & $-2.2\sigma$ & $1.00\pm0.21$  & $-0.9\sigma$\\
\bottomrule
\end{tabular}
\end{table}

\subsection{Astrometric jitter}
\label{sec:simjitter}

The same benchmark with RVs was re-run with extra white noise added to the
astrometry but not to its quoted errors (0.10\,mas). With none added the fitted
jitter is consistent with zero ($0.008^{+0.020}_{-0.006}$\,mas) and every
parameter is within $0.05\sigma$ of a run without the jitter term; with
$0.10$ and $0.30$\,mas added it returns $0.115\pm0.013$ and
$0.312^{+0.022}_{-0.020}$\,mas, and all 16 parameters remain within $2\sigma$ of
the truth while the widths grow. The RV amplitudes are recovered as
$K_1=502.6\pm2.4$ and $K_2=203.1\pm2.4$\,m\,s$^{-1}$ (true 500 and 200).

\subsection{Injection--recovery campaign}
\label{sec:injrec}

To test detection, convergence and the calibration of the posteriors we
injected 180 single companions into the real observing geometry -- epoch times,
scan angles, formal errors and single-star projection -- of five quiet sources
of the DR4 pre-release ($G=9$--14.3, 576--781 epochs, baselines 1733--1902\,d;
their own measurements are not used). The data are
$(\mathbf I-\mathbf H)(\mathbf s+\mathbf n)$ with Gaussian noise $\mathbf n$ at the
formal errors, and are fitted as real data with the default settings (fitted
jitter, isotropic prior, tempered joint fit). The orbits were drawn by Latin
hypercube sampling, in three campaigns of 60 with different seeds, from the
fit's prior restricted to $P=0.1$--2 baselines (log-uniform), $e\le0.9$
(uniform), isotropic orientation, uniform phase, and a total signal-to-noise
ratio $S$ of the projected signal log-uniform from 5 to 100 (which sets $a_0$).
The posteriors are restricted to the same ranges before comparison, so that the
prior equals the injection distribution and the coverage of the credible
intervals must be exact if the fit is correct. A campaign takes 72--91\,min on
9 cores (median 9.3\,min per run).

All 180 runs converged. Tempering is essential here: an earlier campaign with
the untempered sampler converged in only 41 of 60 runs, failing for weak
signals, near-circular orbits -- where a shift of $T_0$ is a rotation that
$A,B,F,G$ absorb, leaving a ring in the posterior -- and periods beyond the
baseline. We re-fitted 23 problem runs (the 19 that failed and 4 that converged
with other problems) and 4 controls with the tempered sampler. All 27 converged,
whereas the untempered sampler had converged only 8 of them, and the tempered
fits took a fifth of the total sampling time. All signals with $S\ge10$ are detected,
and 83\% of those with $S=5$--10; 64\%, 95\%, 91\% and 100\% of the runs
recover the period to 20\% for $S=5$--10, 10--20, 20--50 and 50--100.

Table~\ref{tab:injcal} and Fig.~\ref{fig:injcal} give the calibration over the
173 converged runs with a detection. All parameters are unbiased, and all but
the period are calibrated ($T_0$ slightly conservatively: its intervals are a
little too wide). For the period, 90\% of the truths are
inside the 95\% intervals (17 outliers against 8.7 expected), but the mean
posterior quantile of the truth, $0.48\pm0.02$, and a Kolmogorov--Smirnov test
of its uniformity ($p=0.23$) show no bias. We verified that the prior matches
the injections (reweighting the posteriors to the injection distribution
changes nothing), that the likelihood is exact (the stored single-star gain
uses the same weights $1/\sigma_i^2$), and that the sampler returns the correct
posterior (an independent grid integration, with $A,B,F,G$ marginalised
analytically, reproduces the chains' period quantiles for eight of the worst
runs, and for every outlier the best posterior sample fits better than the
truth). The excess is therefore statistical: it was found in the first campaign
($p=0.03$) and is weaker in the two independent repeats (10 of 115 outside the
95\% intervals against 5.8 expected, $p=0.06$).

Calibration holds on average over the injections, not at each true period.
For true periods within 10\% of the baseline the posterior tends to lie above
the truth (in 6 of 12 runs the truth is at its 0.2--7th percentile, mostly for
$S>20$), balanced by true periods of 1--2 baselines (mean quantile 0.54 against
0.39 for 0.5--1 baselines). We attribute this, without having verified it, to
orbits slightly longer than the baseline reproducing the same projected signal
with more prior volume. \emph{A period close to the baseline may therefore be
overestimated, and its credible interval extends to longer periods.}

Relative astrometry (Sect.~\ref{sec:relast}) was tested with two further
campaigns of 60 injections, with Student-$t$ noise ($\nu=3$) and the
Student-$t$ likelihood: one on the \gaia{} data alone, and one that adds to the
same data sets eight simulated relative positions with $s=a_{\rm rel}/a_0=50$
(Eq.~\ref{eq:relast}), Gaussian errors
of 2\,mas per coordinate and epochs spread over three \gaia{} baselines. All
120 fits converged, and the screen, which uses the \gaia{} data only, detects
the same 53 signals in both. With relative astrometry the posteriors remain
calibrated: averaged over the nine orbital parameters, 71\% and 96\% of the
truths lie inside the 68\% and 95\% intervals (67\% and 94\% without relative astrometry), and 70\%
and 98\% for $s$. Every value is inside the range expected from chance for
53 runs: 55--79\% for the 68\% intervals and 89--100\% for the 95\% intervals. Run by run the credible intervals are narrower, with
median width ratios of 0.27 for $P$, 0.38 for $e$, 0.67 for $a_0$ and 0.64 for
$i$, and the period is within 20\% of the truth in all 53 detected runs,
against 45 without. The gain is largest for the 12 orbits longer than the
\gaia{} baseline, which the relative positions cover in full (0.02 for $P$,
0.10 for $a_0$). Figure~\ref{fig:corner_relast} shows a pair with the typical
gain. The scale $s$ is measured to a median 8\%. On two such
data sets with Gaussian noise ($P=600$ and 3000\,d) \texttt{Octofitter}, fitting
the same along-scan points and relative positions, agrees with
\ExoPlanetLab{} within $0.16\sigma$ in all orbital elements, $s$ and the
parallax, and the widths of the 68\% intervals agree to 4\%. The only difference
is a convention: \texttt{Octofitter} gives the argument of periastron $\omega$
of the companion's orbit about its host, and \ExoPlanetLab{} that of the
photocentre orbit, which differs by $180^\circ$.

\begin{table}
\caption{Calibration of the posteriors in the injection--recovery campaign
(173 converged runs with a detection): fraction of truths inside the central
68\% and 95\% credible intervals, and the median and rms of the normal score
$z=\Phi^{-1}(1-q)$ of the truth's posterior quantile $q$, where $\Phi$ is the
standard normal cumulative distribution function ($z>0$ when the posterior lies
above the truth). The last two rows give the calibrated values and the
binomial 95\% ranges of the coverage for 173 runs.}
\label{tab:injcal}
\centering
\begin{tabular}{lcccc}
\toprule
 & in 68\% & in 95\% & median $z$ & rms $z$\\
\midrule
$P$     & 64\% & 90\% & $+0.09$ & 1.12\\
$e$     & 65\% & 95\% & $+0.02$ & 1.05\\
$T_0$   & 78\% & 97\% & $-0.18$ & 0.87\\
$A$     & 64\% & 91\% & $+0.11$ & 1.08\\
$B$     & 68\% & 95\% & $+0.13$ & 1.02\\
$F$     & 70\% & 94\% & $-0.05$ & 1.00\\
$G$     & 68\% & 95\% & $-0.12$ & 1.02\\
$a_0$   & 63\% & 96\% & $-0.05$ & 1.06\\
$i$     & 69\% & 95\% & $+0.13$ & 1.01\\
\midrule
calibrated & 68\% & 95\% & 0 & 1\\
range      & 61--75\% & 91--98\% & & \\
\bottomrule
\end{tabular}
\end{table}

\begin{figure*}
\centering
\IfFileExists{figures/injection_calibration.pdf}%
  {\includegraphics[width=0.8\textwidth]{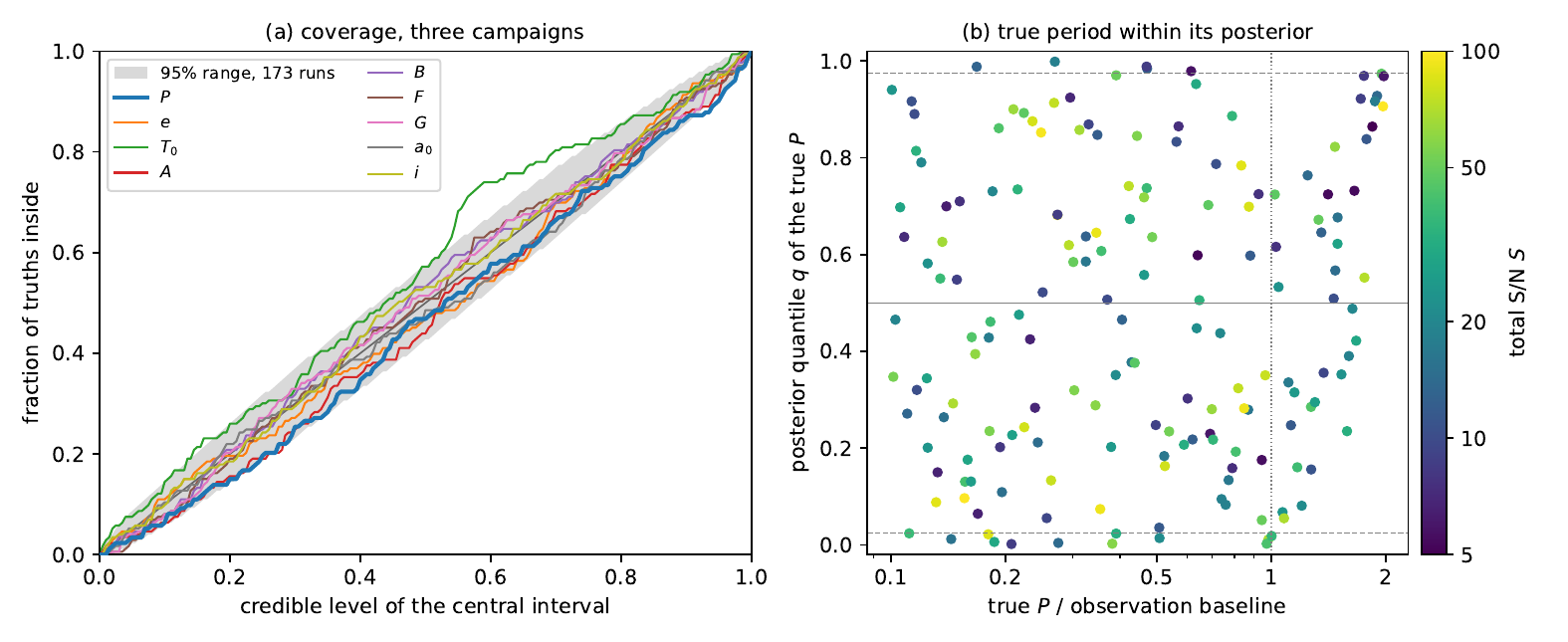}}%
  {\fbox{[injection calibration figure]}}
\caption{Injection--recovery calibration (173 runs). \emph{Left}: fraction of
truths inside the central credible interval against its level, per parameter;
calibrated posteriors follow the diagonal, within the grey 95\% binomial range.
\emph{Right}: posterior quantile of the true period against the true period in
units of the baseline (1733--1902\,d), coloured by the total signal-to-noise ratio $S$; dashed
lines bound the 95\% interval. For true periods just below one baseline the
posterior tends to lie above the truth (points near $q=0$): such periods tend to
be overestimated, while $q$ near 1 means it is underestimated.}
\label{fig:injcal}
\end{figure*}

\subsection{Detection threshold and false positives}
\label{sec:detection}

The tests of this and the following subsections use Student-$t$ noise with
$\nu=3$, like the pre-release data, in the geometry of the same quiet sources,
and single-orbit evidence runs. Thirty data sets without a companion give
$\ln B(1\ \mathrm{vs}\ 0)$ (Sect.~\ref{sec:evidence}) between $-1.44$ and $-0.24$: none is above zero. An
orbit with $P=500$\,d and $e=0.3$ injected at $S=(6, 7, 8, 9, 10)$ with six
noise realisations each is detected ($\ln B>5$) in $(1, 1, 5, 5, 6)$ of them
(Fig.~\ref{fig:detection}a), so half of the companions are found between $S=7$
and 8, and every detected period is within 5\% of the truth. A threshold of 3
would add three detections at $S=7$ without admitting a null, however, 30 nulls do
not exclude a false-positive rate of several per cent.

The screen is the less sensitive stage in such noise. With least squares it
finds 12, 36, 56 and 80\% of the signals at $S=6$, 8, 10 and 12 (50 screens
each); on the Student-$t$ score (Sect.~\ref{sec:screen}) it finds 24, 74, 86
and 100\% (Fig.~\ref{fig:detection}b). The gain costs neither false alarms nor
sensitivity in Gaussian noise. On 300 data sets of pure Student-$t$ noise the
robust screen reports a signal once (the least-squares screen twice), where the
1\% false-alarm threshold allows about three; on 200 data sets of pure Gaussian
noise it reports none; and with a signal at $S=8$ in Gaussian noise both screens
find 49 of 50. For this kind of data one companion is always fitted, so a signal the
screen misses is still recovered by the joint fit from a blind start; the screen
decides how many signals are reported.

\begin{figure*}
\centering
\includegraphics[width=0.8\textwidth]{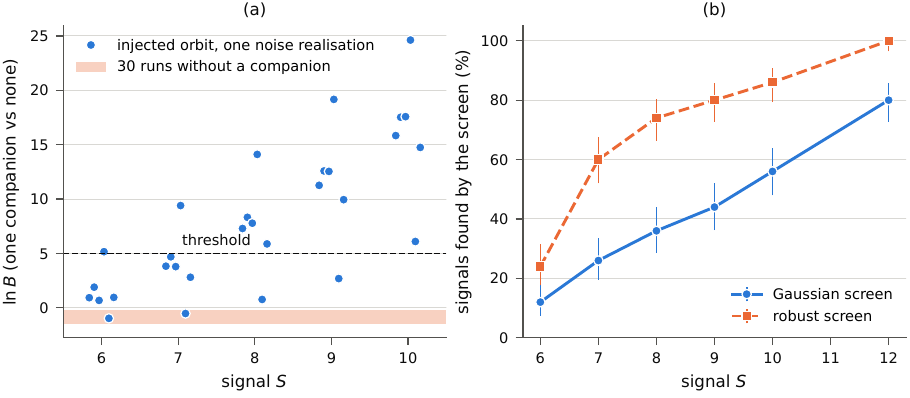}
\caption{Detection in Student-$t$ ($\nu=3$) noise. \emph{(a)}: evidence for one
companion against none for an injected orbit ($P=500$\,d, $e=0.3$), six noise
realisations per signal-to-noise ratio $S$; the band is the range of 30 data
sets without a companion and the dashed line the adopted threshold.
\emph{(b)}: fraction of injected signals found by the periodogram screen with
least squares and with the robust statistic (50 screens per point; 68\%
binomial intervals).}
\label{fig:detection}
\end{figure*}

\subsection{Orientation prior}
\label{sec:priorsens}

For strong signals the orientation prior does not matter; for weak ones it
decides the outcome. At $S=8$ the isotropic prior
gives $\ln B(1\ \mathrm{vs}\ 0)=+15.9$, $+14.5$, $+17.2$ and $+1.7$ in four
tests, and the uniform Thiele--Innes prior $-5.4$, $-7.7$, $-3.2$ and $-18.6$
for the same data, with posteriors that run to $e\approx1$ and an $a_0$ from
twice to thousands of times the true one: its $a_0^3$ volume outweighs a weak
likelihood, as \citet{Baycroft2026} warned. At $S=15$ the two priors give the
same orbit. The effect on the real sources is given in Sect.~\ref{sec:real}.

\subsection{Periods near one year and high eccentricity}
\label{sec:hard}

Orbits with $P=345$, 365.25 and 385\,d, and a 300-d control, were injected at
$S=8$ and 15 with four noise realisations each. At a given $S$ a period of one
year is no harder than 300\,d:
all 16 runs at $S=15$ are detections and 12 of 16 at $S=8$ (the four missed
are at 300, 345 and 385\,d, none at one year), the periods are within
0.2--1.7\% of the truth, and the 28 detected runs are calibrated (the true period inside the 68\%
interval in 68\% of them). The cost of the parallax degeneracy is the amplitude
needed to reach that $S$. 

Orbits with $e=0.8$ and 0.95 are detected in seven of eight runs each (one
missed at $S=8$) and their periods recovered to 1--2\%, but $e$ and $a_0$ come out too small: for $e=0.95$
the posterior median is 0.60--0.76 of the true $e$ and half the true $a_0$, the
truth lying outside the 68\% interval in all seven detected runs. The
eccentricity prior is uniform, so only 5\% of it lies above $e=0.95$, and at
$S=8$--15 the data do not rule out rounder orbits, which then dominate the
posterior. The posteriors are calibrated on average over truths drawn from the
prior (Sect.~\ref{sec:injrec}), not for every fixed truth, and a truth at the
edge of the prior is pulled inwards. A weak, highly eccentric orbit is
therefore reported rounder and smaller than it is.

\subsection{Two companions}
\label{sec:multi}

To test the choice of the number of companions, a second companion was added to
an inner one ($P=300$\,d, $S=15$). It was placed at period ratios of 1.3, 2 and
3, with its own signal $S_2=6$, 10 and 15, in two noise realisations. The 18
cases were fitted with one and two orbits, and four of them also with three
(Table~\ref{tab:multi}). All fits start from the best of several ranked
starting configurations (Sect.~\ref{sec:peel}).

The number of companions is correctly recovered in all 18 cases. At $S_2=6$
the second companion is not accepted ($\ln B$ from $-0.3$ to $+2.8$). At
$S_2=10$ and 15 it always is ($+5.7$ to $+45.5$). A third orbit never is
($-1.9$ to $+1.4$). In all 12 detected cases the two-orbit posterior has a
single period pair, the injected one, with at least 98\% of the samples.

The ranked starts are needed. Started from orbits fitted one after another to
the residuals, the two-orbit fit recovered the second period in only 9 of the
12 detected cases. In the other three, two of them at a period ratio of 1.3,
the posterior held several period pairs, and only 12--48\% of the samples were
on the injected one. In these 16-parameter fits the sampler moves slowly and
stays close to where it starts, which here was a wrong pair.

The one-orbit fit has a similar problem. In one case with two equally strong
companions, at 300 and 600\,d, it found a period of 16.8\,d. This alias fits
the combined signal about as well as either true period. Fits started from
different periods give evidences within 1.6 of each other, and we use the
highest. A second-companion solution is considered reliable when its
posterior has a single period mode or fits from different starts agree. The
test is small as it has one inner orbit, one orientation of the second companion
and only two noise realisations per case. It therefore shows that the number of
companions can be chosen correctly. Its does not how often this succeeds for other
orbits.
\begin{table}
\caption{Two injected companions (inner: $P=300$\,d, $S=15$): evidence for the
second orbit, $\ln B(2\ \mathrm{vs}\ 1)$, for the two noise realisations, and the
period of the second orbit in the two-orbit fit.}
\label{tab:multi}
\centering\footnotesize\setlength{\tabcolsep}{4pt}
\begin{tabular}{@{}llrrll@{}}
\toprule
$P_2$ (d) & $S_2$ & \multicolumn{2}{c}{$\ln B(2\ \mathrm{vs}\ 1)$} & \multicolumn{2}{c}{$P_2$ recovered (d)}\\
\midrule
390 & 6  & $+0.9$  & $+0.3$  & --    & -- \\
    & 10 & $+14.3$ & $+5.7$  & $386^{+5}_{-6}$ & $382\pm6$ \\
    & 15 & $+20.1$ & $+20.8$ & $387\pm4$ & $385\pm4$ \\
600 & 6  & $-0.1$  & $+2.4$  & --    & -- \\
    & 10 & $+13.4$ & $+14.4$ & $595^{+20}_{-15}$ & $579^{+17}_{-15}$ \\
    & 15 & $+16.4$ & $+26.2$ & $598^{+15}_{-12}$ & $586^{+12}_{-11}$ \\
900 & 6  & $-0.3$  & $+2.8$  & --    & -- \\
    & 10 & $+11.0$ & $+17.7$ & $928^{+39}_{-41}$ & $920^{+41}_{-37}$ \\
    & 15 & $+28.9$ & $+45.5$ & $918^{+37}_{-30}$ & $920^{+32}_{-28}$ \\
\bottomrule
\end{tabular}
\tablefoot{One- and two-orbit fits started from the best of several ranked
starting configurations (Sect.~\ref{sec:peel}).}
\end{table}

\subsection{Orbit or acceleration}
\label{sec:orbitaccel}
We examine the period range over which an orbital model is preferred to an 
acceleration model. One companion ($e=0.3$) was injected
with periods of 1 to 4 times the 1733-d time span of the data and $a_0=1$ and
5\,mas. Up to twice the time span the orbit is strongly preferred over the
better of the two acceleration models (with and without jerk;
Fig.~\ref{fig:orbitaccel}). The preference is lost at
about 2.5 time spans for 1\,mas and 3 for 5\,mas, beyond which the acceleration
models have equal or higher evidence and the period posterior is the same
whatever the true period. All twelve are overwhelming detections against the
single star. The comparison separates an orbit from mere curvature. 
For strong signals the jerk term is needed: at 5\,mas and three
time spans the orbit has $\ln B=+66$ against the model with acceleration only,
which alone would wrongly make the orbit look strongly warranted, but $-2.5$
against the one with acceleration and jerk (the value shown in
Fig.~\ref{fig:orbitaccel}), so the curvature is described as well without an
orbit. Both acceleration models must therefore be compared with the orbit.

An inner orbit ($P=300$\,d, $S=15$) with a real outer companion ($a_0=2$\,mas)
shows the use of the terms inside a fit. With one orbit and no acceleration
terms the fit takes the stronger outer signal and loses the inner companion.
For an outer period of 3.5 time spans, acceleration and jerk describe the outer
companion as well as a second orbit does ($\ln\mathcal{Z}=202.2$ against 201.6)
and recover the inner period equally well ($302.5\pm2.2$\,d). The acceleration
alone is not enough. For an outer period of 1.7 time spans the terms are not an
adequate description and two orbits are needed ($\ln B=26$). When the same inner
orbit is injected without an outer companion and fitted with the acceleration
terms, the evidence is only 0.76 lower than without them. The fitted orbit is 
unchanged so including the terms costs little when there is nothing for them to 
describe.

A pure scan-angle signal of 0.04\,mas, injected without an orbit, is detected by
the scan-angle model of Eq.~\eqref{eq:harm} ($\ln B=+6.5$) and not by the orbit
fit ($+0.4$). A real orbit keeps its evidence when the terms of
Eq.~\eqref{eq:harm} with $n=2$ are in the fit: $\ln\mathcal{Z}$ is only 1.2
lower. An orbit does leak into them so a significant scan-angle term by itself 
does not show a signal to be spurious, however, the comparison with the orbit does.

\begin{figure}
\centering
\includegraphics[width=\columnwidth]{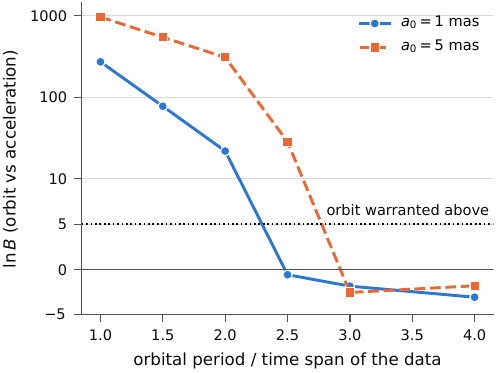}
\caption{Evidence of the one-orbit fit over the better of the acceleration
models (with and without jerk), against the injected period in units of the
time span of the data. The vertical scale is linear below 10 and logarithmic
above.}
\label{fig:orbitaccel}
\end{figure}

\subsection{Upper limits}
\label{sec:limits}

For a source without a detection, we compute an upper limit on the
semi-major axis $a_0$ of a companion as a function of period. This needs no
sampling. At a fixed period, eccentricity and periastron time, the orbit is
linear in its four Thiele--Innes constants. These can therefore be fitted
together with the single-star terms, like the terms of Sect.~\ref{sec:accel}.
The limit is the 95\% point of the posterior of $a_0$. It assumes an
isotropic orientation and a prior uniform in $a_0$, and is marginalised over
eccentricity and phase.

We tested the limits on 400 injected companions for each of two sets. For
circular orbits the true $a_0$ is above the 95\% limit in 2.2\% of the cases.
For eccentric orbits ($e<0.6$) it is above it in 3.5\%. The limits are
therefore slightly conservative.

\section{Application to the \gaia{} DR4 pre-release epoch astrometry}
\label{sec:real}

\begin{table*}
\caption{The three pre-release sources with published orbits: joint fits of the
epoch astrometry and the radial velocities (Student-$t$ noise, posterior medians
and 68\% intervals) next to the published values. $N$ is the number of CCD
measurements, $M_2$ the companion mass from $K$, $i$ and the stellar mass, and
$\varpi$ the parallax corrected for the orbit.}
\label{tab:real}
\centering\footnotesize\setlength{\tabcolsep}{4pt}
\begin{tabular}{@{}lcccccc@{}}
\toprule
 & \multicolumn{2}{c}{Gaia BH3} & \multicolumn{2}{c}{Gaia-4} & \multicolumn{2}{c}{HD\,114762}\\
 & published$^a$ & this work & published$^b$ & this work & published$^c$ & this work\\
\midrule
$N$, time span (d) & & 558, 1861 & & 824, 1804 & & 558, 1798\\
$P$ (d)        & $4253.1\pm98.5$   & $4180.6^{+79.3}_{-74.7}$ & $571.3^{+1.4}_{-1.3}$   & $575.8^{+1.3}_{-1.2}$ & $83.9151\pm0.0030$ & $83.9172^{+0.0008}_{-0.0009}$\\
$e$            & $0.7291\pm0.0048$ & $0.7279\pm0.0038$        & $0.338^{+0.026}_{-0.023}$ & $0.330^{+0.025}_{-0.020}$ & $0.3354\pm0.0048$ & $0.3438\pm0.0011$\\
$a_0$ (mas)    & $27.39\pm0.49$    & $27.07^{+0.40}_{-0.38}$  & $0.279^{+0.016}_{-0.014}$ & $0.249\pm0.008$ & $1.82^{+0.06}_{-0.05}$ & $1.855\pm0.011$\\
$i$ (deg)      & $110.580\pm0.095$ & $110.513\pm0.080$        & $116.9^{+4.2}_{-4.4}$   & $122.9\pm2.6$ & $3.63\pm0.06$ & $6.0^{+3.1}_{-2.7}$\\
$K^d$          & --                & $56.71\pm0.51$           & $368^{+22}_{-18}$       & $330^{+20}_{-22}$ & $612.5\pm3.5$     & $620.4\pm0.8$\\
$M_2^d$        & $32.70\pm0.82$    & $32.5^{+0.9}_{-0.8}$      & $11.8^{+0.7}_{-0.7}$    & $11.5\pm0.9$ & $225\pm14$ & $133^{+124}_{-46}$\\
$\varpi$ (mas) & $1.6933\pm0.0164$ & $1.661\pm0.008$          & 13.628                  & $13.626\pm0.007$ & $25.357\pm0.037$  & $25.30\pm0.01$\\
$\nu$          & --                & $3.0\pm0.4$              & --                      & $8.7^{+3.0}_{-1.8}$ & --                & $2.2\pm0.2$\\
\bottomrule
\end{tabular}
\tablefoot{$^a$ Combined astrometric and spectroscopic solution of
\citet{GaiaBH3}. $^b$ Adopted \gaia{}\,+\,RV solution of \citet{Stefansson2025};
the parallax is the value they adopt. $^c$ $P$, $e$ and $K$: RV orbit of
\citet{Kane2011}, their adopted solution without a linear trend; $a_0$ and parallax: \gaia{} DR3
orbital solution \citep{GaiaArenou2023}; $i$ and $M_2$: \citet{Winn2022}, from
that solution and the RVs.
$^d$ $K$ in km\,s$^{-1}$ and $M_2$ in $M_\odot$ for Gaia BH3; in m\,s$^{-1}$
and $M_{\rm Jup}$ for the other two.}
\end{table*}

\begin{figure*}
\centering
\includegraphics[width=0.8\textwidth]{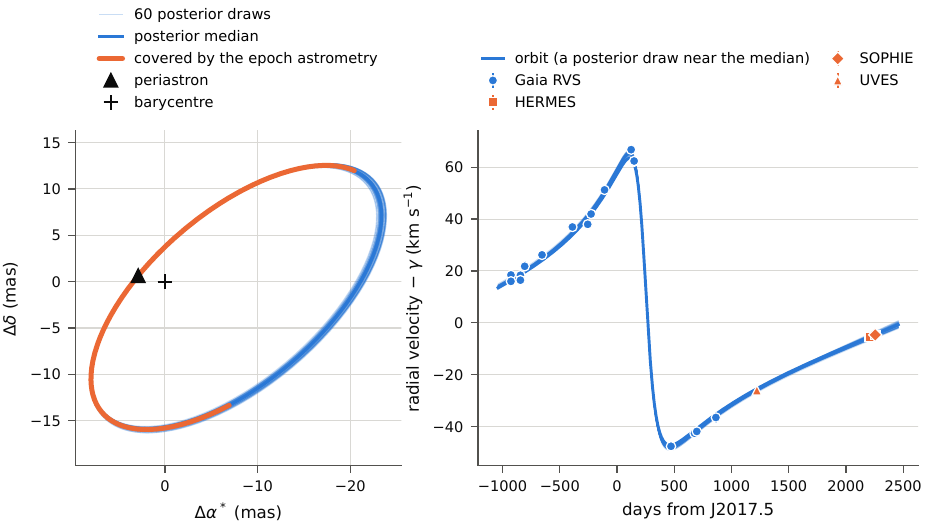}
\caption{Gaia BH3, joint fit of the epoch astrometry and the radial
velocities. \emph{Left}: orbit of the photocentre on the sky (north up, east
to the left) for 60 posterior draws and the posterior median; the part covered
by the 1861 days of epoch astrometry is drawn thicker. \emph{Right}: radial
velocities of \citet{GaiaBH3} with each instrument's fitted offset removed, and
the same posterior draws.}
\label{fig:bh3}
\end{figure*}

The DR4 pre-release (June 2026) contains CCD-level epoch astrometry for 12
sources. We fitted the three with published orbits, with and without RVs
(Table~\ref{tab:real}).

For Gaia BH3 \citep{GaiaBH3} all parameters are within $1.4\sigma$ of both
published solutions. The samplers \texttt{emcee}, \texttt{dynesty} and \texttt{nessai}
agree to $0.14\sigma$ in every parameter (Figs.~\ref{fig:bh3} and
\ref{fig:bh3_corner}). The companion mass agrees with the published one. The
RV amplitude $K$ can also be predicted from the astrometric orbit and the
parallax. The measured $K$ agrees with this prediction for $\varpi=1.696$\,mas,
the parallax of the published combined solution. For our orbit-corrected
parallax it is $2.0\sigma$ below the prediction, and the purely astrometric
mass is correspondingly 6\% higher. No parallax zero-point is applied in the
fit. Subtracting the \gaia{} DR3 zero-point of \citet{Lindegren2021zp} for this
source, $-0.035$\,mas, from our parallax gives $1.696\pm0.008$\,mas. Our results
are therefore consistent with a zero-point of that size in the pre-release
astrometry.

In the joint fit $K$ is free. Tying it to the astrometric orbit
(Sect.~\ref{sec:rv}) tests whether the data agree with a dark companion, and
gives one mass instead of an astrometric and an RV-based one. For Gaia BH3 the
orbit then moves by $0.6\sigma$. The parallax becomes $1.668\pm0.007$\,mas and
the single mass $33.9\pm0.5\,M_\odot$. For none of the three sources do the
data contradict a dark companion.

Fitted to the epoch data of \citet{GaiaBH3} (their Table~B.1), the method
reproduces their astrometric solution to better than $0.7\sigma$ in every
parameter. The remaining differences on the pre-release data come from the
differences between the two reductions of the same CCD measurements. After
removing the single-star terms, their positions differ by $0.11\mas$ rms per
CCD, comparable to the formal errors. Adding these measured differences to the
data of \citet{GaiaBH3}, reassigned at random among transits, moves the
solution by 1.2--$1.8\sigma$. This contains every difference we find.

For HD\,114762 the astrometry gives a nearly face-on orbit. The data have a
large excess noise from heavy-tailed CCD outliers: with the formal errors,
$\chi^2/\mathrm{dof}=9.9$ (Sect.~\ref{sec:jitter}). With a single white jitter,
the astrometry-only period and eccentricity ($83.850\pm0.038$\,d,
$0.311\pm0.029$) lie $1.8\sigma$ and $1.1\sigma$ from the RV orbit. The
inclination is then bounded only from above ($i<20$\,deg, $M_2>39\,M_{\rm Jup}$,
95\%). With the Student-$t$ model the astrometry alone gives
$P=83.915\pm0.013$\,d and $e=0.3445\pm0.010$, within $0.15\sigma$ of the RV orbit.
The joint fit then bounds the inclination: $i<11$\,deg and $M_2>71\,M_{\rm Jup}$
(95\%, isotropic orientation prior). 90\% of the posterior is above the
hydrogen-burning limit.

HD\,114762 is the only one of the three sources whose posterior depends on the
orientation prior (Sect.~\ref{sec:priorsens}). With RVs its inclination is
$6.0^{+3.1}_{-2.7}$\,deg under the isotropic prior and $8.4^{+2.6}_{-2.8}$\,deg
under the uniform Thiele--Innes prior. Gaia-4 and Gaia BH3 have the same
posteriors under both. The companion is most likely an M dwarf.
\citet{Kiefer2019} first concluded this from the DR1 excess noise.
\citet{Winn2022} combined the \gaia{} DR3 orbital solution with the RVs and
found $i=3.63\pm0.06$\,deg and $M_2=0.215\pm0.013\,M_\odot$
($225\pm14\,M_{\rm Jup}$). Our mass in Table~\ref{tab:real},
$133^{+124}_{-46}\,M_{\rm Jup}$, is lower and much wider. With $K$ free, the
inclination comes from the astrometric orbit alone. For a nearly face-on orbit
this bounds it only from above, and $M_2\propto1/\sin i$ follows the prior.
With $K$ tied to the astrometric orbit through the parallax
\citep[][Eq.~12]{PourbaixJorissen2000}, which the data allow, the RVs fix the
inclination as well. We then find $i\approx3.5$\,deg and $M_2\approx240\,M_{\rm Jup}$, close to the values of \citet{Winn2022}.

For Gaia-4 \citep{Stefansson2025} the eccentricity, inclination and $T_p$ agree
with the published solution, but the period and orientation differ by
2--$3\sigma$. The published solution rests on DR3 data. The period differs by
$3.3\sigma$ of the published width, or $2.5\sigma$ of the two widths combined.
The errors of Gaia-4 are nearly Gaussian, and the differences are the same with
the formal errors, a Gaussian jitter and the Student-$t$ model. We also held
the published orbit fixed and computed $\chi^2$ against the epoch data, for
$N_{\rm obs}=824$ measurements. With no orbit, $\chi^2$ is $2869$. With the
orbit projected through the single-star fit (Sect.~\ref{sec:projection}), it
is $1227$. Without the projection it is $3087$, worse than no orbit at all.
For Gaia BH3 ($N_{\rm obs}=558$), $\chi^2$ is $3.7\times10^6$ with no orbit,
$2016$ with the projected orbit and $6.7\times10^6$ without the projection. The projection is therefore essential when a published
orbit is compared with the epoch data.

\begin{figure}
\centering
\includegraphics[width=\columnwidth]{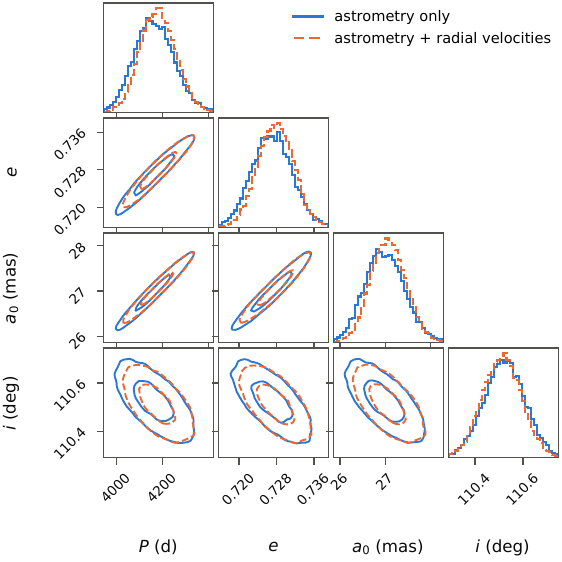}
\caption{Gaia BH3: posterior of the period, eccentricity, photocentre
semi-major axis and inclination from the astrometry alone (solid) and with the
radial velocities (dashed). Contours enclose 39\% and 86\% of the probability.}
\label{fig:bh3_corner}
\end{figure}

With the formal errors, the orbit fits give $\chi^2/\mathrm{dof}=1.4$ for
Gaia-4, 3.1 for Gaia BH3 and 9.9 for HD\,114762. The excess sits within transits and is
heavy-tailed (Sect.~\ref{sec:jitter}). It is not an error common to the CCDs of
a transit, and part of it belongs to particular gates
(Sect.~\ref{sec:features}). Its origin could be calibration, the source (e.g.\
the photocentre of a partly resolved companion) or other unmodelled effects;
we cannot identify it. Refitting Gaia BH3 with an error factor per gate moves
its orbit by less than $0.4\sigma$.

With the Student-$t$ model, the orbital values of Gaia BH3 change by less than
their uncertainties, compared with a Gaussian model with a fitted jitter
($0.116$\,mas). The uncertainties shrink by about a quarter: $\pm77$ against
$\pm102$\,d in $P$, and $\pm0.0038$ against $\pm0.0051$ in $e$. They are now
somewhat narrower than the published ones ($\pm98.5$\,d, $\pm0.0048$), which use
Gaussian errors.

The evidences select one companion for each of the three sources. Against no
companion, $\ln B=+370$, $+1082$ and $+2429$ (Gaia-4, HD\,114762, Gaia BH3, in
that order throughout). For a second companion, $\ln B$ is $-1.9$ for Gaia-4
and $-1.6$ for Gaia BH3. For HD\,114762 the tempered \texttt{emcee} fit with two orbits did
not reach a stable evidence in 200\,000 steps. Its second orbit collapses to
the floor of the amplitude prior. The nested samplers give $\ln B=-0.3\pm0.4$
(\texttt{nessai}) and $-1.4\pm0.6$ (\texttt{dynesty}).

The one-orbit evidence of HD\,114762 needed the stability test of
Sect.~\ref{sec:evidence}. A run stopped by the autocorrelation time of the cold chain gave
$\ln\mathcal{Z}=100$, too low by 14.
Longer runs give $\ln\mathcal{Z}=114$, and so do \texttt{nessai} and
\texttt{dynesty}. For Gaia BH3 with RVs the two nested samplers disagree by 4,
so we use the tempered evidence there.

The orbit is strongly preferred over the simpler models. Against the better
of the two acceleration models, $\ln B=+362$, $+1053$ and $+1722$. Against the
better of the two scan-angle models, $\ln B=+365$, $+1023$ and $+2304$. We also added the acceleration
terms or the scan-angle terms (Sect.~\ref{sec:accel}) to the
orbit fit. The evidence then drops by 1--4. The fitted accelerations and
scan-angle amplitudes are all consistent with zero, and the orbits are
unchanged. None of the three sources shows curvature beyond its orbit or a
scan-angle signal. The nine other pre-release sources show no companion,
acceleration or scan-angle term above the threshold. For them we computed
95\% upper limits on $a_0$ (Sect.~\ref{sec:limits}). The limit varies with
period. Its median over periods up to the time span of the data is
0.03--0.06\,mas at $G=9$--14 and 0.6--1.7\,mas at $G=19$--20.5.

Three other codes were run on the same measurements. \texttt{Octofitter}
\citep{Thompson2023} reduces each transit to the median of its CCD
measurements. Fitted to the astrometry alone, it agrees with our fits within
$0.8\sigma$ on every parameter of the three sources. \texttt{kepmodel} (maximum
likelihood, Gaussian noise) agrees within $0.4\sigma$ in every parameter for
Gaia-4 and Gaia BH3. It also reproduces our Gaussian-model fit of HD\,114762.
\texttt{kima} (Student-$t$ likelihood, priors matched to ours) agrees within
$1.1\sigma$ for Gaia-4, including the tail parameter $\nu$. Its sampler did not
converge for the two strong signals.

Recovering the single-star terms for each posterior draw
(Sect.~\ref{sec:projection}) gives the orbit-corrected parallaxes of
Table~\ref{tab:real}. For Gaia BH3 the parallax is $1.2\sigma$ below that of the
published astrometric solution ($1.6747\pm0.0094$\,mas). It is $1.8\sigma$ below
that of the combined solution before a zero-point correction, and $0.1\sigma$
above it with the DR3 zero-point of \citet{Lindegren2021zp}. For Gaia-4 it
equals the value adopted by \citet{Stefansson2025}. For HD\,114762 it is
$1.6\sigma$ below the parallax of the DR3 orbital solution,
$25.357\pm0.037$\,mas. It is $3.4\%$ below the DR3 single-star parallax
(26.198\,mas, RUWE 3.2), which ignores the 84-d orbit. The pseudo-colour prior has a negligible effect on the projection
(Sect.~\ref{sec:projection}). With the fitted orbits, the projected $\chi^2$
and that of a joint fit without the prior differ by less than $2\times10^{-5}$
for all three sources.

\section{Discussion and conclusions}
\label{sec:discussion}
\label{sec:conclusions}

\subsection{Summary of results}

\ExoPlanetLab{} fits Keplerian orbits to \gaia{} epoch astrometry, alone or
with radial velocities and relative astrometry, with one likelihood and five
samplers. The principal results are the following.
\begin{enumerate}
\item The model orbit must be compared with the epoch residuals only after
removing the part that the pipeline's single-star fit absorbs. Without this
projection, a published orbit fits the DR4 epoch data worse than no orbit at
all. The single-star parameters are recovered for every posterior draw.
\item The excess noise of the DR4 pre-release data is heavy-tailed at the CCD
level. A Student-$t$ likelihood with robustly refitted single-star terms
describes it. For HD\,114762 it brings the astrometric period and eccentricity
onto the RV orbit. A robust version of the periodogram screen detects at
$S=7$--8 what least squares detects at $S=10$--12.
\item Three pre-release sources have published orbits. For Gaia BH3 all
parameters are within $1.4\sigma$ of both published solutions. For Gaia-4 the
period and orientation differ by 2--$3\sigma$ from the published solution,
which rests on DR3 data. For HD\,114762 the astrometry bounds the inclination,
$i<11$\,deg (95\%). The companion is therefore not a planet but most likely an
M dwarf, as \citet{Winn2022} found from the DR3 orbit and the RVs. With $K$
free its mass follows the orientation prior and lies below theirs. With $K$
tied to the astrometric orbit through the parallax we find $i\approx3.5$\,deg
and $M_2\approx240\,M_{\rm Jup}$, close to their values. The
orbit-corrected parallaxes of the three sources are within $1.8\sigma$ of the
published ones.
\item The samplers agree with each other, and with \texttt{Octofitter},
\texttt{kepmodel} and (for Gaia-4) \texttt{kima} on the same data. In 180
injections, the posteriors of the 173 runs with a detection are unbiased. All
parameters but the period are calibrated; 90\% of the true periods lie within
their 95\% credible intervals. Highly eccentric orbits at low signal-to-noise ratio are reported
rounder and smaller than they are. A uniform Thiele--Innes prior loses weak
signals.
\item Evidences of fits with a fixed number of companions select that number.
There is no false detection in 30 data sets without a companion, and half of
the companions are detected at $S=7$--8. With two injected companions the count
is right in all 18 cases. The orbit of the second companion depends on where
the fit starts. The fit therefore starts from the best of several ranked
candidates. The evidence of a strong signal
needs a temperature ladder adapted to equilibrated chains and a stability test.
Without them it was low by 10--14 for HD\,114762.
\item Acceleration and scan-angle models show three things. An orbit is
warranted up to periods of 2--3 times the time span of the data. An inner
orbit survives a wide companion when acceleration and jerk are fitted with it.
None of the three pre-release companions is a scan-angle artefact or carries a
second signal.
\item Relative astrometry of a resolved companion enters with one parameter,
the ratio $s$ of its relative orbit to the photocentre orbit. With the
parallax, it gives the total mass without a stellar model. In paired
injections, eight relative positions leave the posteriors calibrated. They
narrow the credible intervals to 0.27 ($P$), 0.38 ($e$), 0.67 ($a_0$) and 0.64
($i$) of their widths, and far more for periods beyond the time span of the
data. \texttt{Octofitter} agrees within $0.2\sigma$ on the same simulated data.
\end{enumerate}

\subsection{Limitations and scope of the validation}
\label{sec:features}
\label{sec:tests}

The excess noise of the pre-release data is heavy-tailed, and its origin is not
identified. Part of it depends on the gate. An error factor per gate, however,
explains little of the heavy tail. The number of degrees of freedom $\nu$ of
the Student-$t$ likelihood (Sect.~\ref{sec:jitter}), which is small for heavy
tails, rises only from 2.9 to 4.0 for Gaia BH3 and from 2.2 to 3.0 for
HD\,114762. The Student-$t$ likelihood treats the CCD measurements as
independent. No term common to the CCDs of a transit is detected.

The log-uniform $a_0$ prior extends far below the detection limit. An
undetectable extra companion, acceleration or scan-angle term therefore lowers
$\ln\mathcal{Z}$ by only 1--4. A term is added only when it raises
$\ln\mathcal{Z}$ by more than 5, and in our tests no spurious term was added.
A negative $\ln B$ of $-1$ to $-4$, however, only says that the data do not
need the term. It does not rule the term out.

With two or more orbits the tempered sampler does not move between period
modes. The result therefore depends on the ranked starts
(Sect.~\ref{sec:peel}). The posterior can have several modes, for example the
true pair of periods and an alias pair. A single run then finds only one of
them, and the evidences of separate runs are needed to weigh them.

A fit of one companion with its evidence takes about one core-hour for a source
with 600 measurements. The time grows in proportion to the number of
measurements, and faster with the number of companions.

The validation has limits of its own. The injections use the geometry of five
sources at $G=9$--14. Fainter sources were only screened and given upper
limits. The real orbits are those of three sources with strong signals and
published solutions. The tests with two companions and of the transition from
orbit to acceleration are small, with one orbit shape and one or two noise
realisations per case. Relative astrometry was tested on simulated data only.

\subsection{Outlook}

\ExoPlanetLab{} can be applied to the DR4 epoch astrometry as released, alone
or with radial velocities and relative astrometry. The first test after the
release is a comparison with the orbital solutions of a larger sample, such as
the \gaia{} DR3 non-single stars. Each source is fitted independently, so a
large sample is spread over as many cores as are available. At about one
core-hour per source, $10^5$ sources take about four days on a cluster with
1000 cores. For the full DR4 sample the main open question is therefore the
origin of the excess noise.

\begin{acknowledgements}
This work has made use of data from the European Space Agency (ESA) mission
\gaia{} (\url{https://www.cosmos.esa.int/gaia}), processed by the \gaia{} Data
Processing and Analysis Consortium (DPAC, \url{https://www.cosmos.esa.int/web/gaia/dpac/consortium}). 
This work was financially supported directly by the Swedish National Space Agency (SNSA/Rymdstyrelsen). 
The code and its documentation have been developed by the author working with Claude AI (Opus 5.5 and 
earlier Sonnet 5.5) using a Human Intelligence (HI)--AI iterative procedure. Many hundreds of HI-AI 
iterations have been necessary during development and testing and the author takes full responsibility 
for any remaining errors. 
\end{acknowledgements}

\paragraph{Code availability.} \ExoPlanetLab{} is available at
\url{https://github.com/hobbsSe/ExoPlanetLab} and archived at Zenodo,
\url{https://doi.org/10.5281/zenodo.24101622} (version 1.0.0), under the GPL-3.0 licence.
The code comes with a user guide (\texttt{ExoPlanetLab.pdf}). It describes the
model, the priors, the samplers and every setting, and gives worked examples
and exercises.

\IfFileExists{bibtex/aa.bst}{\bibliographystyle{bibtex/aa}}{%
  \IfFileExists{aa.bst}{\bibliographystyle{aa}}{\bibliographystyle{plainnat}}}
\bibliography{ExoPlanetLabPaper}

\begin{appendix}
\renewcommand{\floatpagefraction}{0.3}
\section{Corner plots}
\label{app:corner}

Figures~\ref{fig:corner_sim}--\ref{fig:corner_relast} show posteriors of the
period, eccentricity, photocentre semi-major axis and inclination in pairs of
fits to the same astrometry, in the style of Fig.~\ref{fig:bh3_corner}:
contours enclose 39\% and 86\% of the probability, and the histograms are
normalised.
\begin{figure*}
\centering
\includegraphics[width=1.2\columnwidth]{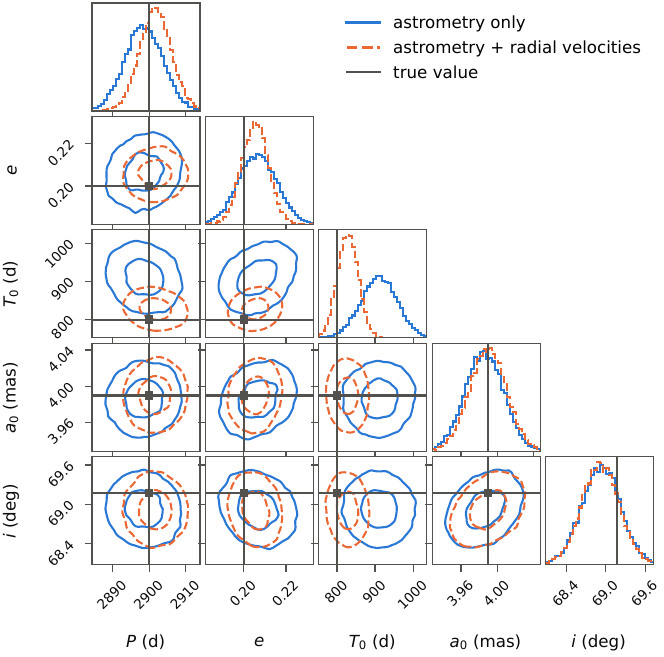}
\caption{Outer planet of the simulated 2:1 system (Sect.~\ref{sec:sims},
Table~\ref{tab:sim}), observed for 1.2 orbits, from the astrometry alone (solid)
and with the radial velocities (dashed); thin dark lines and squares mark the injected values (true values).
Without RVs the periastron time $T_0$ lies along a ridge that leaves its median
$2$--$3\sigma$ from the truth; the RVs remove it.}
\label{fig:corner_sim}
\end{figure*}

\begin{figure*}
\centering
\includegraphics[width=1.2\columnwidth]{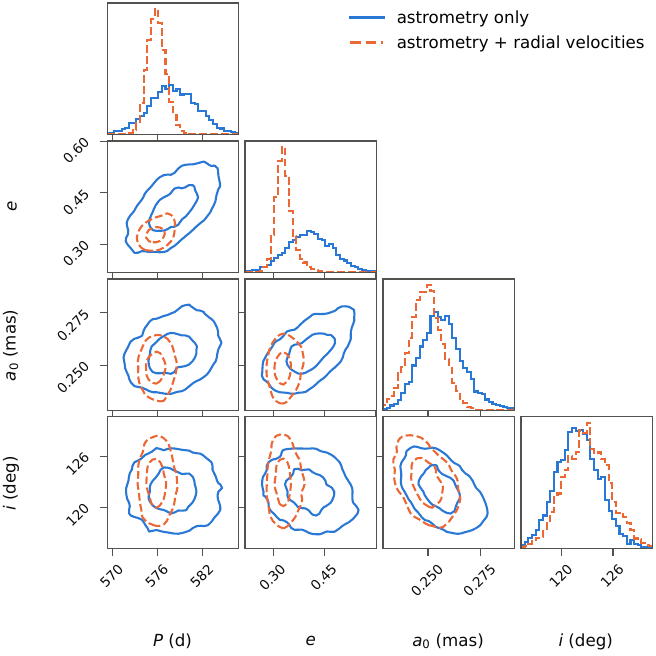}
\caption{Gaia-4, from the astrometry alone (solid) and with the radial
velocities of \citet{Stefansson2025} (dashed).}
\label{fig:corner_gaia4}
\end{figure*}

\begin{figure*}
\centering
\includegraphics[width=1.2\columnwidth]{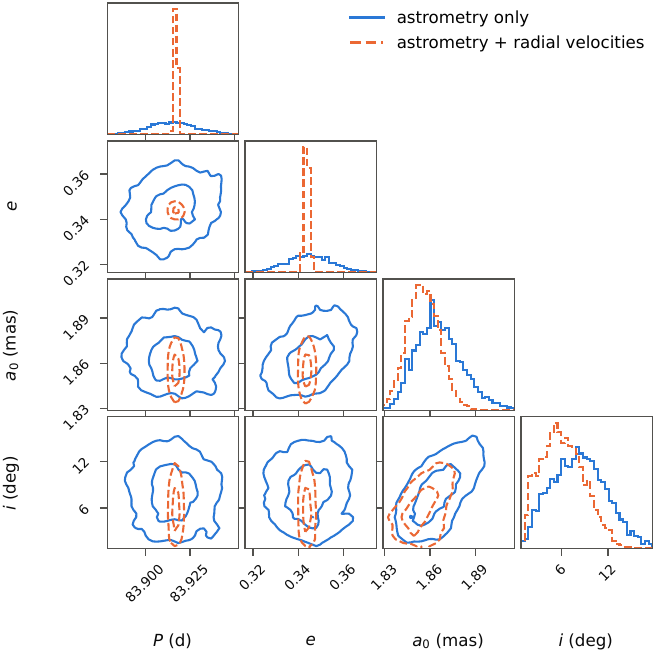}
\caption{HD\,114762, from the astrometry alone (solid) and with the radial
velocities (dashed), which fix the period and eccentricity; the orbit is close
to face-on.}
\label{fig:corner_hd114762}
\end{figure*}

\begin{figure*}
\centering
\includegraphics[width=1.2\columnwidth]{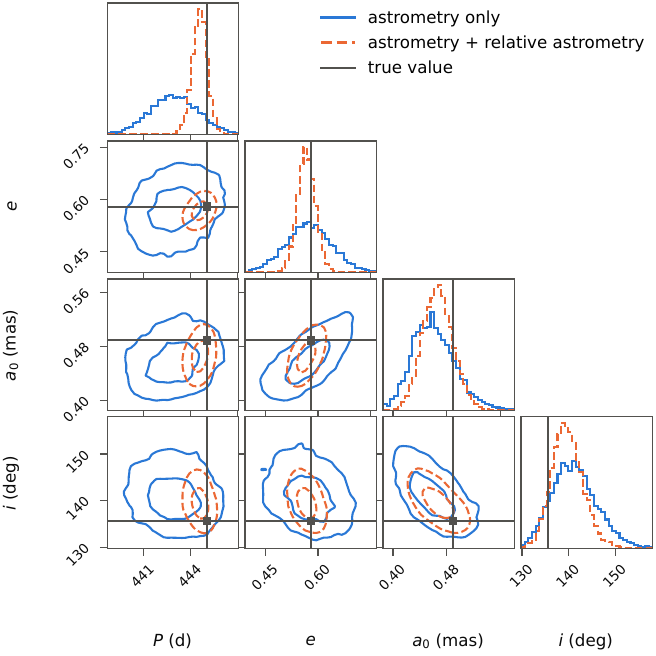}
\caption{An injected companion ($P=445$\,d, $e=0.58$, $S=40$;
Sect.~\ref{sec:injrec}) from the astrometry alone (solid) and with eight
simulated relative positions (dashed); thin dark lines and squares mark the injected values (true values).
This pair has the typical gain of the campaign: the 68\% intervals narrow to
0.31 ($P$), 0.39 ($e$), 0.71 ($a_0$) and 0.68 ($i$) of their widths. For periods
beyond the time span of the astrometry the gain is far larger (0.01--0.03 in
$P$).}
\label{fig:corner_relast}
\end{figure*}
\end{appendix}

\end{document}